\documentclass[preprint,showpacs,aps,pra]{revtex4}
\usepackage{epsfig}
\usepackage{psfrag}
\usepackage{graphicx}
\usepackage{titlesec}
\usepackage{graphics}
\usepackage{amsmath}
\usepackage{color}
\usepackage[utf8]{inputenc}
\usepackage[T1]{fontenc}
\usepackage[title]{appendix}

\usepackage{tikz} 
\begin{document}

\title{CP-violating non-linear electrodynamics and corrections to blackbody radiation thermal laws}

\author{L.P.R. Ospedal} \email{leoopr@cbpf.br}
\affiliation{Centro Brasileiro de Pesquisas F\'isicas, Rua Dr. Xavier Sigaud 150, CEP 22290-180, Rio de Janeiro, RJ, Brazil}

\author{R. Turcati} \email{turcati@cbpf.br}
\affiliation{Centro Brasileiro de Pesquisas F\'isicas, Rua Dr. Xavier Sigaud 150, CEP 22290-180, Rio de Janeiro, RJ, Brazil}
\affiliation{Departamento de Física, Universidade Federal de Santa Catarina, Campus Trindade, 88040-900, Florianópolis, Brazil}

\author{S.B. Duarte} \email{sbd@cbpf.br}
\affiliation{Centro Brasileiro de Pesquisas F\'isicas,  Rua Dr. Xavier Sigaud 150, CEP 22290-180, Rio de Janeiro, RJ, Brazil}
\begin{abstract}

Motivated by ideas from effective field theories, we conduct a twofold investigation in CP-violating non-linear electrodynamics models. On the one hand, features related to field equations and wave propagation in the presence of a magnetic background field are evaluated. Interestingly, we show that the CP-violating term in our framework induces a bi-anisotropic behavior of the vacuum. On the other hand, blackbody radiation thermal laws in this situation are studied. Here, we provide a general formalism for photons at low temperature, namely, with $k_{B}T\ll{m_{e}c^{2}}$. The deviations from standard values of the thermodynamic quantities, such as energy, pressure, entropy, and specific heat, are discussed. In addition, we investigate the modifications in the Planck distribution and the Stefan-Boltzmann law induced by a non-trivial CP-violating parameter.

\end{abstract}
%
\maketitle

\section{Introduction} \label{sec_1}

Back to 1956, the pioneering work of Lee and Yang on
the possible parity (P) symmetry breaking in the Electroweak interaction (EW) \cite{Lee_Yang_PR56} and, subsequently, the experimental confirmation by the group led by Wu \cite{Wu_PR57} pointed out  the relevance of discrete symmetries for a better understanding of fundamental interactions. At this stage, the 
parity symmetry remains preserved by the strong nuclear and electromagnetic interactions.  After that, the CP symmetry, namely, the combination of charge conjugation (C) and parity transformation, emerged as a candidate to be the true symmetry between matter and antimatter, but the group led by Cronin and Fitch discovered that this symmetry would also be broken by EW in decay of neutral kaon \cite{Cronin_PRL64}. Some years later, violations of CP were observed in many experiments involving meson and baryon decays \cite{KTev_PRL99,NA48_PLB99,BABAR_PRL01,Belle_PRL01,LHCb_PRL13,LHCb_PRL19,LHCb_nature25}, as well as possible recent indications of its breakdown in the leptonic sector \cite{T2K_nature20,Denton_PRL21,Amir_PRD21,NOva_PRL24}. Nowadays, it is well-known that the Standard Model (SM) of elementary particles  accommodates a CP-violating contribution in the quark sector with the Cabibbo–Kobayashi–Maskawa (CKM) matrix. In addition, a small CP-violating term is allowed in the context Quantum Chromodynamics (QCD), with the presence of the so-called $\Theta-$term. It is also important to emphasize that the combined CPT transformations, where T denotes time reversal transformation, is a fundamental symmetry for any local Quantum Field Theory, which is Lorentz invariant and has a hermitian Hamiltonian. Therefore, a violation of CP automatically implies into a violation of T, such that CPT is preserved. Furthermore, as pointed out by Sakharov \cite{Sakharov_SP91}, the violation of CP appears as one of the conditions for the emergence of matter-antimatter asymmetry in the primordial universe.  If the contribution of the Standard Model to CP breaking is too small to explain this asymmetry, then we will need to search for new physics Beyond the Standard Model (BSM), either by introducing new interactions, gauge groups or fields. For more details on CP violation in the context of SM and proposals BSM, we highlight the Refs. \cite{BOOK_Branco,BOOK_Ikaros09,BOOK_Ikaros21}. Within these notable research lines and motivations, we will be particularly concerned with CP-violating contributions to non-linear electrodynamics. As we shall see, one may gain further insight into new physics BSM by probing CP-violating terms in the photonic sector, which is typically associated with effective field theories.


A lot of efforts has been devoted to investigate general aspects of non-linear electrodynamics, which sometimes includes  CP-violating contributions. We highlight the seminal works of 1970 by Bialynicka-Birula and Bialynicki-Birula \cite{Birula_PRD70}, and independently by Boillat \cite{Boillat_JMP70}, where the authors discussed the properties of photon propagation in the presence of an external electromagnetic field.  Furthermore, as we shall see in the next Section \eqref{sec_2}, it is well-known that the field equations of non-linear electrodynamics can be written in terms of constitutive relations involving the vectors ${\bf D}$ and ${\bf H}$ that resemble the description of non-linear optical media. With this in mind, the correspondent optical phenomena such as birefringences, photon splitting, second harmonic generation, wave mixing, and other optical properties were investigated in Refs. \cite{Novello_PRD00,Lorenci_PLB00,Battesti_RPP13,Rizzo_EPJD06,Robertson_PRA19,Liao_EPJ08,Querts_EPJC23,Spallicci_PLB24}. Mention should be made that the predictions of non-linear electrodynamics are the subject of intense research and the correspondent parameters can be tested or constrained by many experiments, which encompasses investigations from optical effects, photon-photon scattering, electric and magnetic dipole moments, high intensity lasers, energy levels of the hydrogen atom, among others \cite{Fouche_PRD16,Fedotov_PR23,Cruz_EPJC25}. For a recent review and remarkable progress on non-linear electrodynamics, we point out the work \cite{Sorokin_FP22} and references therein. 



At this point, it is appropriate to discuss some possible generations of CP-violating non-linear electrodynamics. The main motivations are associated with effective field theories.  As expected, within the scenario of a purely Quantum Electrodynamics (QED), which is invariant under the discrete symmetries C, P, and T separately, there is no contribution to CP-violating terms even if taking into account quantum corrections in any order of perturbative expansion. However, in the context of the Standard Model, it is worthwhile mentioning that a CP-violating photon self-interaction can be generated from an effective contribution of the strong nuclear interaction, namely, from the so-called $\Theta-$term of QCD, but this contribution is quite small \cite{Faccioli_PRD09}. Therefore, the appearance of non-trivial contributions to CP-violating terms could indicate possible effects of new physics BSM. On the other hand, some phenomenological investigations in the photonic sector could be useful for constraining the parameters related to proposals BSM. Initially, let us consider the CP-violating terms associated with new couplings. For instance, we point out that the inclusion of both vector and pseudo-vector couplings between fermions and photon can generate an effective CP-violation Euler-Heisenberg-like electrodynamics \cite{Yamashita_PETP17,Ghasemkhani_EPJ23}. Analogously, a similar model may be obtained from an effective theory related to non-minimal electromagnetic tensor and pseudo-tensor couplings, although these contributions are quite limited due to the current measurements of magnetic and electric dipole moments \cite{Ghasemkhani_EPJ22,Chupp_RMP19,PDG}. Other possible sources are the introduction of dark matter candidates coupled to the photonic sector, such that the corresponding effective theories lead  to CP-violating self-photon interactions. The most popular approach is the incorporation of the axion field \cite{Gorghetto_JHEP21,Paradisi_PRD21,Paradisi_JHEP24}. Another viewpoint is the inclusion of a new abelian dark sector $U_{Y'}(1)$ coupled to the Electroweak sector $SU_L(2) \times U_Y(1)$ of the Standard Model \cite{Fan_PETP2018}, which introduces new scalar, spinor, and vector fields so that the low-energy effective theory also yields a CP-violating non-linear electrodynamics. At last but not least, by taking into account quantum corrections in the Standard Model, it is possible to give rise to some effective theories with anomalous couplings. In particular, non-linear corrections in the hypercharge sector $U_Y(1)$ can induce anomalous neutral couplings between photon and $Z^0$ boson, as well as contributions to CP-violating photonic terms. This previous scenario is also motivated by effective low-energy models in the context of String Theories (see, for example, the work \cite{Remmen_JHEP19} and references therein). 

With these ideas and motivations regarding photon propagation in mind, it would be worthwhile to consider the implications for the thermal laws governing blackbody radiation. Indeed, electromagnetic radiation in thermal equilibrium has played a fundamental role in modern physics, including the description of cavity radiation, which led to energy quantization in quantum systems and the birth of quantum mechanics, as well as in the characterization of the cosmic microwave background (CMB), among other phenomena. In a blackbody, electromagnetic frequencies are excited by thermalization, and Planck's law accurately describes its properties at thermal equilibrium. On the other hand, corrections to photon propagation should induce small deviations from the standard blackbody radiation laws, which, for instance, modify the Stefan-Boltzmann law and thermodynamic potentials. In certain scenarios $-$ such as  extra dimensions \cite{Ramaton}, massive photons \cite{Torres_PRA85,Moreira_BJP46,Saiyan_CBAO71}, Lorentz symmetry violation \cite{Casana_PRD08,Casana_PRD10,Anacleto:2018wlj,Petrov_EPJC21}, non-linear electrodynamics \cite{NiauAkmansoy:2013sxs,Turcati_PRD23,Turcati_JPA24} and quantum gravity proposals \cite{Filho_ADP22,Turcati_CQG24} $-$ deviations from the standard blackbody radiation predictions within Maxwell's theory, including the Stefan-Boltzmann law and thermodynamic potentials, have been investigated in recent years, opening new frontiers in modern physics.




The main purpose of this paper is to further elaborate about modified dispersion relations in the presence of a magnetic background field and the violation of CP in photon self-interactions, as well as examining the implications to the blackbody radiation thermal laws. This work is organized as follows. Initially, in Section \ref{sec_2}, we obtain the dispersion relations for a general CP-violating non-linear electrodynamics in the presence of a magnetic background field. In addition, we also present an overview of possible scenarios in which CP-violating contributions may be generated. After that, in Section \ref{sec_3}, our approach aims to investigate an effective low-energy CP-violating model and its contributions to thermodynamic properties such as energy density, pressure, entropy, and specific heat. Furthermore, we derive the correspondent corrections to the Planck distribution and Rayleigh-Jeans law, as well as we get the modified Stefan-Boltzmann law. Some phenomenological implications are also discussed. Finally, in Section \ref{sec_concl}, we exhibit our concluding comments and perspectives. Throughout this work, we adopt the SI units, where $\varepsilon_0$ and $\mu_0$ denote the vacuum permittivity and permeability, respectively. Furthermore, $c= 1/\sqrt{\varepsilon_0 \, \mu_0}$ corresponds to the speed of light, while $k_B$ stands for the Boltzmann constant. The metric convention is given by $\eta^{\mu \nu} = \textrm{diag}(+,-,-,-)$. 


\section{The model under consideration} \label{sec_2}

In this Section, we present the modified field equations of a photon propagation in the presence of a uniform and constant magnetic background field. We also take the opportunity to discuss the constitutive relations and comparisons between the vacuum of non-linear electrodynamics and material media. After that, the modified dispersion relations and group velocities are obtained. Finally, we exhibit an overview of CP-violating non-linear electrodynamics and useful results for particular models.

\subsection{Photon propagation in a magnetic background field} \label{subsec_2A}

We start off with the general non-linear Lagrangian density described by
\begin{equation} \label{defi_L}
\mathcal{L} = \frac{1}{\mu_0} \, L_{nl} - \rho \, \phi + {\bf j} \cdot {\bf A} \, ,
\end{equation}
where $\rho$ and ${\bf j}$ correspond to the charge and current densities, respectively, while $\phi$ and ${\bf A}$ denote the electromagnetic potentials such that  ${\bf E} = - \nabla \phi - \partial {\bf A} / \partial t$ and ${\bf B} = \nabla \times {\bf A}$. The  action is given by $S = \int dt \,  d^3 {\bf x} \, \mathcal{L}$. In addition, we assume that $L_{nl} = L_{nl}(\mathcal{F}, \mathcal{G})$, with $\mathcal{F}$ and $\mathcal{G}$ being the Lorentz and gauge invariants defined as follows
\begin{eqnarray}
    \mathcal{F} &\equiv& - \frac{1}{4} F_{\mu \nu}^2 = \frac{1}{2} \left( \frac{ {\bf E}^2 }{c^2} - {\bf B}^2 \right) 
    \, , \label{inv_F} \\
    \mathcal{G} &\equiv& - \frac{1}{4} \, F_{\mu \nu} \widetilde{F}^{\mu \nu} = \frac{ {\bf E} \cdot {\bf B} }{c} 
    \, , \label{inv_G}
\end{eqnarray} 
where $F_{\mu \nu} = \partial_\mu A_\nu - \partial_\nu A_\mu$ denotes the field strength tensor of the potential $A_\mu = (\phi/c \, , -{\bf A})$. Additionally, $\partial_\mu = (\frac{1}{c} \, \partial_t , \partial_i \equiv \nabla_i )$ and $\widetilde{F}^{\mu \nu} \equiv \epsilon^{\mu \nu \alpha \beta} \, F_{\alpha \beta}/2$ is the dual tensor, with $\epsilon^{\mu \nu \alpha \beta}$ being the Levi-Civita symbol in Minkowski spacetime $(\epsilon^{0123}=1)$. According to these definitions, we have the following components
\begin{equation} \label{F_comp}
    F_{0i} = {\bf E}_i /c \, \; , \, \; 
    F_{ij} = - \epsilon_{ijk} \, {\bf B}_k \, \; , \, \;
    \widetilde{F}_{0i} = {\bf B}_i \, \; , \, \;
    \widetilde{F}_{ij} = \epsilon_{ijk} \, {\bf E}_k /c \, ,
\end{equation}
where $\epsilon_{ijk}$ stands for the euclidean Levi-Civita symbol $(\epsilon^{0ijk} \equiv \epsilon_{ijk} \, \;\textrm{and} \, \; \epsilon_{123} =1)$. Although, the covariant formulation will not be used in this and the next subsection, we take this opportunity to establish our conventions. This will be only relevant in the subsection \ref{subsec_2C}, where we make some comparisons with other models in the literature.

In what follows and for our purposes, the non-covariant description is sufficient. Hence, we follow a similar approach of Ref. \cite{Haas_POP23}. The field equations related to the Lagrangian density \eqref{defi_L} are given by
\begin{eqnarray}
\nabla \cdot {\bf D} &=& \rho \label{eq_D} \, , \\
\nabla \times {\bf H} &=& {\bf j} + \frac{\partial {\bf D}}{\partial t} \label{eq_H} \, ,
\end{eqnarray}
where we introduced the auxiliary fields
\begin{eqnarray}
{\bf D} & = & \frac{\partial L_{nl} }{\partial \mathcal{F} } \, \varepsilon_0 \, {\bf E} + 
\frac{\partial L_{nl} }{\partial \mathcal{G} } \, c \, \varepsilon_0 \, {\bf B} \label{def_D} \, , \\
{\bf H} & = & \frac{\partial L_{nl} }{\partial \mathcal{F} } \, \frac{ {\bf B} }{\mu_0} - 
\frac{\partial L_{nl} }{\partial \mathcal{G} } \, \frac{ {\bf E} }{\mu_0 \, c} \label{def_H} \, .
\end{eqnarray}
For Maxwell's electrodynamics, we have $ L_{nl} = \mathcal{F} $, which leads to ${\bf D} = \varepsilon_0 \, {\bf E} $ and ${\bf H} =  {\bf B} / \mu_0 $ such that Eqs. \eqref{eq_D} and \eqref{eq_H} recover the usual inhomogeneous Maxwell equations.  

Here, we point out that the homogeneous equations remain unchanged, 
\begin{equation}
\nabla \times {\bf E} = - \frac{\partial {\bf B}}{\partial t} \, \; , \, \;
\nabla \cdot{\bf B} = 0 \, . \label{homo_eqs}  
\end{equation}

Let us now assume electromagnetic fields around a  magnetic background  field ${\bf B}_0$, in a such way that $ {\bf E} = {\bf  e} $ and $ {\bf B} = {\bf B}_0 + {\bf b} $, where ${\bf e}$ and ${\bf b}$ are small perturbations. 
Having these in mind, it is possible to expand Eqs. \eqref{def_D} and \eqref{def_H} in the linear perturbations, which lead to 
\begin{eqnarray}
 {\bf D} &  = & c_1 \, \varepsilon_0 \, {\bf e} + c_2 \, \varepsilon_0 \, c \left( {\bf B}_0 + {\bf b} \right)
+ d_2 \, \varepsilon_0 \, {\bf B}_0 \left( {\bf B}_0 \cdot {\bf e} \right)
- d_3 \, \varepsilon_0 \, c \, {\bf B}_0 \left( {\bf B}_0 \cdot {\bf b} \right)
\label{lin_D} \, , \\
 {\bf H} & = & \frac{c_1}{\mu_0} \, \left( {\bf B}_0 + {\bf b} \right) 
- \frac{c_2}{\mu_0 c} \, {\bf e}  
- \frac{d_1}{\mu_0} \, {\bf B}_0 \left( {\bf B}_0 \cdot {\bf b} \right) 
+ \frac{d_3}{\mu_0 c} \, {\bf B}_0 \left( {\bf B}_0 \cdot {\bf e} \right)
\label{lin_H} \, ,
\end{eqnarray}
with the coefficients $c_1 , c_2 , d_1 , d_2$ and $d_3$ being evaluated at the magnetic background field, namely,
\begin{eqnarray} 
c_{1} &=& \left. \frac{\partial L_{nl}}{\partial{\cal F}}\right|_{{\bf B}_0}
\; \, , \; \,
c_{2} = \left. \frac{\partial L_{nl}}{\partial{\cal G}}\right|_{{\bf B}_0}
\; \, , \; \, \label{coef_c} \\
d_{1} &=& \left. \frac{\partial^2 L_{nl}}{\partial{\cal F}^2}\right|_{{\bf B}_0}
\; \, , \; \,  
d_{2} = \left. \frac{\partial^2 L_{nl}}{\partial{\cal G}^2}\right|_{{\bf B}_0}
\; \, , \; \,
d_{3} = \left. \frac{\partial^2 L_{nl}}{\partial{\cal F}\partial{\cal G}}\right|_{{\bf B}_0}
\; \, . \label{coef_d}
\end{eqnarray}
For the Maxwell electrodynamics,  $ L_{nl} = \mathcal{F} $, we  have $c_1=1$ and $c_2 = d_1 = d_2 = d_3 = 0$.

Under parity $(P)$ transformation (${\bf x}' \rightarrow - {\bf x}$), we have that $ {\bf E}' \rightarrow  - {\bf E} $ and 
$ {\bf B}' \rightarrow   {\bf B} $, or equivalently, the invariants $\mathcal{F}' \rightarrow  \mathcal{F} $ and $\mathcal{G}' \rightarrow  - \mathcal{G} $. For this reason, terms with odd powers in $\mathcal{G}$ will break the parity symmetry. For example, contributions in $L_{nl} = L_{nl}(\mathcal{F}, \mathcal{G})$ with $\mathcal{F G}, \mathcal{F}^2 \mathcal{G}, \mathcal{G}^3$ and so on. Thus, from Eqs. \eqref{lin_D} and \eqref{lin_H}, one can conclude that non-trivial contributions of $c_2$ and $d_3$ will lead to parity violating terms. As charge conjugation $(C)$ symmetry is preserved by the invariants $\mathcal{F}$ and $\mathcal{G}$ (since $A_C^\mu = - A^\mu$), it is quite common to refer to the  models with $c_2 \neq 0$ and $d_3 \neq 0$ as CP-violating non-linear electrodynamics.

At this stage, we also consider the magnetic background field to be uniform and constant. Thus, from Eq. \eqref{homo_eqs} and the aforementioned assumptions, we get the homogeneous equations for the perturbation fields
\begin{eqnarray}
\nabla \times {\bf e} &=& - \frac{\partial \, {\bf b} }{\partial t}  \, , \label{Faraday_e_b} \\ 
\nabla \cdot{\bf b} &=& 0 \label{Gauss_b}  \, .
\end{eqnarray}
By using these homogeneous equations, one can eliminate the terms with the coefficient $c_2$ in Eqs. \eqref{eq_D} and \eqref{eq_H}, which allows us to redefine the auxiliary fields as
\begin{eqnarray}
 {\bf D} &  = & c_1 \, \varepsilon_0 \, {\bf e} 
+ d_2 \, \varepsilon_0 \, {\bf B}_0 \left( {\bf B}_0 \cdot {\bf e} \right)
- d_3 \, \varepsilon_0 \, c \, {\bf B}_0 \left( {\bf B}_0 \cdot {\bf b} \right)
\label{lin_D_2} \, , \\
 {\bf H} & = & \frac{c_1}{\mu_0} \,  {\bf b} 
- \frac{d_1}{\mu_0} \, {\bf B}_0 \left( {\bf B}_0 \cdot {\bf b} \right) 
+ \frac{d_3}{\mu_0 c} \, {\bf B}_0 \left( {\bf B}_0 \cdot {\bf e} \right)
\label{lin_H_2} \, ,
\end{eqnarray}
Hence, only the coefficients $c_1, d_1, d_2$ and $d_3$ will contribute to the modified field equations.

Before proceeding, it is interesting to rewrite the auxiliary fields in terms of the following matrix components 
\begin{eqnarray}
    {\bf D}_i &=& \varepsilon_{ij} \, {\bf e}_j + \zeta_{ij} \, {\bf b}_j \, , \label{D_comp} \\
    {\bf H}_i &=& (\mu^{-1})_{ij} \, {\bf b}_j + \xi_{ij} \, {\bf e}_j \, , \label{H_comp}
\end{eqnarray}
where we identified
\begin{eqnarray}
    \varepsilon_{ij} & = & c_1 \, \varepsilon_0 \, \delta_{ij} + d_2 \, \varepsilon_0 \, {\bf B}_{0i} \, {\bf B}_{0j} 
    \, , \label{varepsilon_def} \\
    \zeta_{ij} & = & - d_3 \, \varepsilon_0 \, c \, {\bf B}_{0i} \, {\bf B}_{0j} \, , \label{zeta_def} \\
    (\mu^{-1})_{ij} & = & \frac{c_1}{\mu_0} \, \delta_{ij} - \frac{d_1}{\mu_0} \, {\bf B}_{0i} \, {\bf B}_{0j} 
    \, , \label{mu_def} \\
    \xi_{ij} & = & \frac{d_3}{c \, \mu_0} \, {\bf B}_{0i} \, {\bf B}_{0j} \label{xi_def} \, ,
\end{eqnarray}
with $\delta_{ij}$ being the Kronecker delta. From Eqs. \eqref{D_comp} and \eqref{H_comp}, we can establish some comparisons between the vacuum of non-linear electrodynamics and the constitutive relations of material media. Firstly, we mention that $ \varepsilon_{ij}$ and $ (\mu^{-1})_{ij}$ correspond to the permittivity and permeability tensors, respectively. We  highlight that the coefficient $c_1$ only appears in the isotropic contributions, i.e., with the terms proportional to $\delta_{ij}$. In the presence of non-trivial CP-violating term $(d_3 \neq 0)$, we also obtain the tensors $\zeta_{ij}$ and $\xi_{ij}$ such that the vacuum behaves like a bi-anisotropic material, namely, both ${\bf e}$ and ${\bf b}$ can generate contributions to the magnetization and polarization vectors in different directions according to the components of the magnetic background field. Otherwise, when $d_3=0$, the vacuum plays the role of an anisotropic material, with ${\bf B}_0$ and the coefficients $d_1$ and $d_2$ being responsible for the anisotropy terms.

In what follows, we consider the non-linear electrodynamics in vacuum ($\rho=0$ and ${\bf j} = { \bf 0}$). Therefore, with the auxiliary fields \eqref{lin_D_2} and \eqref{lin_H_2} in Eqs. \eqref{eq_D} and \eqref{eq_H}, we arrive at 
\begin{equation}
    \nabla \cdot {\bf e} + \frac{d_2}{c_1} \, {\bf B}_0 \cdot \nabla \left( {\bf B}_0 \cdot {\bf e} \right) - \frac{c \, d_3}{c_1} \, {\bf B}_0 \cdot \nabla \left( {\bf B}_0 \cdot {\bf b} \right) = 0 \, , \label{mod_Gauss_law}
\end{equation}
\begin{eqnarray}
  \nabla \times {\bf b}  &+& \frac{d_1}{c_1} \, {\bf B}_0 \times \nabla \left( {\bf B}_0 \cdot {\bf b} \right) - 
  \frac{d_3}{c \, c_1} \, {\bf B}_0 \times \nabla \left( {\bf B}_0 \cdot {\bf e} \right) = \nonumber \\
  &=& \frac{1}{c^2} \, \frac{\partial \, {\bf e}}{\partial t}  + \frac{d_2}{c^2 \, c_1} \, {\bf B}_0 \, \frac{\partial}{\partial t} \left( {\bf B}_0 \cdot {\bf e} \right)
  - \frac{d_3}{c \, c_1} \, {\bf B}_0 \, \frac{\partial}{\partial t} \left( {\bf B}_0 \cdot {\bf b} \right) \, . \label{mod_A-M_law}
\end{eqnarray}

The set of field equations  \eqref{Faraday_e_b}, \eqref{Gauss_b}, \eqref{mod_Gauss_law} and \eqref{mod_A-M_law} 
describes the propagation of electromagnetic fields of a photon
in the presence of a constant and uniform magnetic background field.

\subsection{Modified dispersion relations} \label{subsec_2B}

In order to obtain the modified dispersion relations of the previous set of equations, we assume plane wave solutions
\begin{equation} \label{plane-wave}
    {\bf e} ({\bf x},t) = {\bf e}_0 \, e^{i ({\bf k} \cdot {\bf x} - \omega \, t)} \, \; , \, \; 
    {\bf b} ({\bf x},t)  = {\bf b}_0 \, e^{i ({\bf k} \cdot {\bf x} - \omega \, t)} \, ,
\end{equation}
where ${\bf k}$ and $\omega$ denote the wave vector and frequency, respectively. From Eqs. \eqref{Faraday_e_b} and \eqref{Gauss_b}, we promptly get $ {\bf b}_0 = ({\bf k} \times {\bf e}_0) / \omega$ and $ {\bf k} \cdot {\bf b}_0 = 0 $. By using these conditions in the modified Gauss equation \eqref{mod_Gauss_law}, one arrives at the following condition
\begin{equation} \label{Gauss_k}
    {\bf k} \cdot {\bf e}_0 = - \frac{d_2}{c_1} \, 
    \left( {\bf B}_0 \cdot {\bf k} \right) 
    \left( {\bf B}_0 \cdot {\bf e}_0 \right) +
    \frac{c \, d_3}{c_1} \, 
    \left( {\bf B}_0 \cdot {\bf k} \right) 
    \left( {\bf B}_0 \times {\bf k} \right) 
    \cdot \frac{{\bf e}_0}{\omega} \, .
\end{equation}
Furthermore, together with the modified Ampere-Maxwell equation \eqref{mod_A-M_law}, we obtain
\begin{eqnarray} \label{Ampere_Maxwell_k}
     \left( \frac{\omega^2}{c^2} - {\bf k}^2 \right) {\bf e}_0 
     &+& \left[ \frac{c \, d_3}{c_1 \, \omega} 
     \left( {\bf B}_0 \cdot {\bf k} \right) {\bf k} + 
    \frac{d_1}{c_1}  \left( {\bf B}_0 \times {\bf k} \right) 
    - \frac{\omega \, d_3}{c \, c_1} \, {\bf B}_0 \right] 
     \left( {\bf B}_0 \times {\bf k} \right) \cdot {\bf e}_0 + 
     \nonumber \\
     &+&\left[ \frac{\omega^2 \, d_2}{c^2 \, c_1} \, {\bf B}_0 
     - \frac{d_2}{c_1}  \left( {\bf B}_0 \cdot {\bf k} \right) {\bf k} - \frac{\omega \, d_3}{c \, c_1} \,
     \left( {\bf B}_0 \times {\bf k} \right) \right] 
     \left( {\bf B}_0 \cdot {\bf e}_0 \right)= 0 \, .
\end{eqnarray}
At this point, we adopt a similar methodology of Ref. \cite{Neves_PRD21}, namely, we rewrite Eq. \eqref{Ampere_Maxwell_k} into the form $M_{ij} \, {\bf e}_{0j} = 0 $, with the matrix components
\begin{equation} \label{M_def}
   M_{ij} = \alpha \, \delta_{ij} \,  + {\bf u}_{1,i} \, {\bf u}_{2,j} + {\bf u}_{3,i} \, {\bf u}_{4,j} \, , 
\end{equation}
%
%
%
%
%
where we find that
\begin{eqnarray}    
     \alpha & = &  \frac{\omega^2}{c^2} - {\bf k}^2 \, , \\
     {\bf u}_{1,i} & = & \frac{d_1}{c_1} ({\bf B}_0 \times {\bf k})_i - \frac{\omega \, d_3}{c \, c_1} \, {\bf B}_{0i}  
     + \frac{c \, d_3}{c_1 \, \omega}  \left( {\bf B}_0 \cdot {\bf k} \right) {\bf k}_i \, , \\
     {\bf u}_{2,j} &=& \left( {\bf B}_0 \times {\bf k} \right)_j 
     \, , \\
    {\bf u}_{3,i} &=& \frac{\omega^2 \, d_2}{c^2 \, c_1} \, {\bf B}_{0i} 
    - \frac{\omega \, d_3}{c \, c_1} \left( {\bf B}_0 \times {\bf k} \right)_i - \frac{d_2}{c_1} \left( {\bf B}_0 \cdot {\bf k} \right) {\bf k}_i \, , \\
    {\bf u}_{4,j} &=& {\bf B}_{0j} \, \, .    
\end{eqnarray}

%
%

 We now use the result that the determinant of a matrix $M \, \;(3 \times 3)$  can be expressed in terms of products involving the traces of the matrices $M$, $M^2$, and $M^3$, namely, 
\begin{equation} \label{det_m_trace}
    \det (M) = \frac{1}{6}  \left( \textrm{tr} M\right)^3 - \frac{1}{2} \left(  \textrm{tr} M \right) \textrm{tr} \left( M^2 \right) + \frac{1}{3} \, \textrm{tr} \left( M^3 \right) \, .
\end{equation}
Therefore, it is possible to show that the determinant of the matrix $M$ is given by
\begin{equation} \label{det_M}
    \det (M) = \alpha \left[ 
    (\alpha + {\bf u}_1 \cdot {\bf u}_2) 
    (\alpha + {\bf u}_3 \cdot {\bf u}_4)
    - ({\bf u}_1 \cdot {\bf u_4}) ({\bf u}_2 \cdot {\bf u}_3)    
    \right] \, .
\end{equation}
%
%
%
%
%
The non-trivial solutions of $M_{ij} \, {\bf e}_{0j} = 0 $ are obtained by imposing that $\det(M) = 0$. Firstly, we have the condition $\alpha =0$, which leads to the usual photon dispersion relation $ \omega^2 = c^2  \, {\bf k}^2 $. In addition, the other solutions arise from the following condition $(\alpha + {\bf u}_1 \cdot {\bf u}_2) (\alpha + {\bf u}_3 \cdot {\bf u}_4)- ({\bf u}_1 \cdot {\bf u}_4) ({\bf u}_2 \cdot {\bf u}_3)  = 0$, or equivalently, from the polynomial equation
%
%
%
%
\begin{equation} \label{pol_w_eq}
    P \, \frac{\omega^4}{c^4} + Q \, \frac{\omega^2}{c^2} + R = 0 \, ,
\end{equation}
with the definitions
\begin{eqnarray}
    P &=& 1 + \frac{d_2}{c_1} \, {\bf B}_0^2  \, , \\
    Q &=& - 2 {\bf k}^2 - \frac{d_2}{c_1} \left[ \left( {\bf B}_0 \cdot {\bf k} \right)^2 + {\bf B}_0^2 \, {\bf k}^2 \right] + \frac{(d_1 d_2 - d^2_3)}{c_1^2} \, {\bf B}_0^2
     \left( {\bf B}_0 \times {\bf k} \right)^2 \, , \\
    R &=& {\bf k}^4 + \frac{d_2}{c_1} \, {\bf k}^2 
    \left( {\bf B}_0 \cdot {\bf k} \right)^2 
    - \frac{d_1}{c_1} \, {\bf k}^2 
    \left( {\bf B}_0 \times {\bf k} \right)^2 
    -  \frac{(d_1 d_2 - d^2_3)}{c_1^2} \, 
     \left( {\bf B}_0 \cdot {\bf k} \right)^2
      \left( {\bf B}_0 \times {\bf k} \right)^2 \, .
\end{eqnarray}
By solving this polynomial equation, we arrive at the  frequencies
\begin{eqnarray} \label{DR}
    \omega_{(\pm)} & =& \frac{c}{\sqrt{2 \left( 1 + d_2 {\bf B}_0^2 /c_1 \right) }} \, \left\{
    2 {\bf k}^2 \left( 1+ \frac{d_2 {\bf B}_0^2 }{c_1}\right)
    - \frac{ \left( {\bf B}_0 \times {\bf k} \right)^2}{c_1} \, \left[ d_1 + d_2 + \frac{\left(d_1 d_2 - d^2_3 \right)}{c_1} \, {\bf B}_0^2 \right]  + \right. \nonumber \\
    & \pm & \left. \frac{ \left( {\bf B}_0 \times {\bf k} \right)^2}{c_1} \, \sqrt{ \left[ d_1 - d_2 + \frac{\left( d_1 d_2 - d^2_3 \right)}{c_1} \, {\bf B}_0^2 \right]^2 + 4 \, d_3^2} \right\}^{1/2} \, .
\end{eqnarray}

These dispersion relations describe the non-trivial photon propagating modes in the presence of magnetic background field, where both ${\bf B}_0$ and ${\bf k}$ have arbitrary directions. In the particular situations with ${\bf B}_0 = {\bf 0}$ or ${\bf B}_0  \parallel  {\bf k}$, we recover the usual dispersion relation $\omega^2 = c^2 \, {\bf k}^2$.  

Here, it is worthy to observe that, for the particular case with parity invariance $(d_3=0)$, we recover the well-known results in the literature \cite{Neves_PRD21}, namely, 
\begin{eqnarray} 
    \omega_{(-)} \, \rightarrow \, \omega_1 & \equiv & c \,  k \, \sqrt{ 1 - \frac{d_1}{c_1} \, ( {\bf B}_0 \times {\bf \hat{k}} )^2 } \, , \label{w1} \\
    \omega_{(+)} \, \rightarrow \, \omega_2 & \equiv & c \, k \, \sqrt{ 1 - \frac{d_2 \, ({\bf B}_0 \times {\bf \hat{k}} )^2}{c_1 + d_2 \, {\bf B}_0^2} } 
    \, , \label{w2}
\end{eqnarray}
where we defined $ k = |{\bf k}|$ and ${\bf \hat{k}} \equiv {\bf k} /k$.

 Moreover, we would like to point out that the methodology used in this work to obtain the dispersion relations differs from the standard one \cite{Birula_PRD70,Boillat_JMP70}. However, our approach  has the advantage of being easily adapted to other situations in which non-linear electrodynamics are coupled with other dynamical or external fields. For example, we quote a case of coupling with the axion field \cite{Paixao_JHEP22}, as well as in the context of 
 Lorentz symmetry violation \cite{Paixao_JHEP24}. It should be noted that the situation described here, i.e., non-linear electrodynamics with CP-violating symmetry, had not yet been addressed from the viewpoint of this methodology. Bearing in mind possible generalizations, which will be properly discussed in the Section of our Concluding Comments and Perspectives \ref{sec_concl}, we believe it would be useful to present this alternative and interesting approach.

Next, we pass to describe the  group velocities of the frequencies \eqref{DR}. Remembering the usual definition, ${\bf v}_{g \, i} = \partial \omega/ \partial {\bf k}_i $, we obtain
\begin{eqnarray} \label{vg}
    {\bf v}_g \bigg|_{\omega_{(\pm)}} &=& 
    \frac{c}{\sqrt{2 \left( 1 + d_2 {\bf B}_0^2 /c_1 \right) }} \, \Biggl\{ 2 + \frac{d_2}{c_1} \, \left[ ({\bf B}_0 \cdot {\bf \hat{k}})^2 + {\bf B}_0^2 \right] 
    - \frac{d_1}{c_1} \, ({\bf B}_0 \times {\bf \hat{k}})^2  \nonumber + \\
    &-& \frac{\left( d_1 d_2 - d^2_3 \right)}{c_1^2} \, {\bf B}_0^2 \, ({\bf B}_0 \times {\bf \hat{k}})^2 \pm 
    \frac{({\bf B}_0 \times {\bf \hat{k}})^2}{c_1} 
    \sqrt{\left[ d_1 - d_2 + \frac{\left( d_1 d_2 - d^2_3 \right)}{c_1} \, {\bf B}_0^2 \right]^2 + 4 \, d_3^2} \, \Biggr\}^{-1/2} \times \nonumber \\
    &\times& \Biggl\{ 2 {\bf \hat{k}} + \frac{d_2}{c_1} \, \left[ {\bf B}_0 ({\bf B}_0 \cdot {\bf \hat{k}})+ {\bf B}_0^2 \, {\bf \hat{k}} \right] + 
    \frac{d_1}{c_1} \, {\bf B}_0 \times ({\bf B}_0 \times {\bf \hat{k}})
    + \frac{\left( d_1 d_2 - d^2_3 \right)}{c_1^2} \, {\bf B}_0^2 \, [{\bf B}_0 \times ({\bf B}_0 \times {\bf \hat{k}})] \nonumber \\
    &\mp& \frac{{\bf B}_0 \times ({\bf B}_0 \times {\bf \hat{k}})}{c_1} \sqrt{\left[ d_1 - d_2 + \frac{\left( d_1 d_2 - d^2_3 \right)}{c_1} \, {\bf B}_0^2 \right]^2 + 4 \, d_3^2} \Biggr\} \, .
\end{eqnarray}

Finally, it should be mentioned that these group velocities satisfy the following condition
\begin{equation} \label{vg_iso}
    {\bf v}_g \bigg|_{\omega_{(\pm)}}  \cdot {\bf \hat{k}} = 
    \frac{\omega_{(\pm)}}{k} =  v_p \bigg|_{\omega_{(\pm)}} \, ,
\end{equation}
where ${\bf v}_p $ denotes the phase velocity and $ v_p \equiv |{\bf v}_p|$, i.e., the radial component of group velocity coincides the phase velocity module. From the results in Eqs. \eqref{DR}, \eqref{vg} and \eqref{vg_iso}, we are ready to proceed with the calculations of modified blackbody radiation thermal laws. We shall return to this point in the next Section \ref{sec_3}.

\subsection{An overview of CP-violating non-linear electrodynamics} \label{subsec_2C}

In this subsection, it is appropriate to go into detail about the possible generation of CP-violating non-linear electrodynamics. As already mentioned in the Introduction \ref{sec_1}, the main motivations are related to effective field theories. In this sense, one can investigate the following expansion
\begin{equation} \label{gen_L_eff}
    L_{nl} (\mathcal{F}, \mathcal{G}) = \sum_{i, \,j} \, a_{ij} \, \mathcal{F}^i \, \mathcal{G}^j \, ,
\end{equation}
with $a_{ij}$ being some constants.

Initially, let us consider the model
\begin{equation} \label{L_CP}
    L^{CP}_{nl} = \mathcal{F} + \xi_1 \, \mathcal{F}^2 + \xi_2 \, \mathcal{G}^2 + \xi_3 \, \mathcal{F} \, \mathcal{G} \, , 
\end{equation}
where $\xi_1, \xi_2$ and $\xi_3$ are also constants with the same dimension as the inverse square of the magnetic field. It is not necessary to include the term with only $ \mathcal{G}$ because this will be a total derivative (topological contribution) in the action. 

From the definitions \eqref{coef_c} and \eqref{coef_d} applied to Eq. \eqref{L_CP}, we obtain 
\begin{equation} \label{coef_c_d_LCP}
    c_1 = 1 - \xi_1 {\bf B}_0^2 \, \, , \, \;
    d_1 = 2 \, \xi_1 \, \, , \, \;
    d_2 = 2 \, \xi_2 \, \, , \, \;
    d_3 = \xi_3 \, .
\end{equation}
Thus, only the coefficient $c_1$ depends on the magnetic background field, while the others are constants.

For example, the well-known low-energy Euler-Heisenberg (EH) electrodynamics corresponds to
\begin{equation} \label{EH_coeff}
    \xi_1^{EH} = \frac{8}{45} \frac{\alpha^2}{m_e^4} \frac{\hbar^3 \varepsilon_0}{c^3} 
    \, \; , \, \; 
    \xi_2^{EH} = \frac{14}{45} \frac{\alpha^2}{m_e^4} \frac{\hbar^3 \varepsilon_0}{c^3}    
    \, \; , \, \;
    \xi^{EH}_3=0 \, \; , \, \;
\end{equation}
which takes into account the first-order (one-loop) contributions of Quantum Electrodynamics (QED) to light-by-light interactions \cite{Heisenberg_Euler,Dunne_04}. Here, $\hbar$ and $m_e$ denote the reduced Planck constant and electron mass, respectively, while $ \alpha \equiv e^2/(4\pi \varepsilon_0 \hbar c) \approx 1/137 $ is the fine structure constant. By using the definition of the Schwinger-Sauter critical magnetic field, $B_c \equiv m_e^2 c^2/(e\hbar) \approx 4,41 \times 10^{9} \, T$, one can promptly verify that $\xi_1^{EH} = 2 \alpha/(45 \, \pi B_c^2) \approx 5,28 \times 10^{-24} \, T^{-2}$ and $\xi_2^{EH} = 7 \alpha/(90 \, \pi B_c^2) \approx 9,24 \times 10^{-24} \, T^{-2}$. Since QED is invariant under C, P, and T separately, it is expected that $\xi^{EH}_3= 0$ in all orders in the perturbative expansion.  However, within the Standard Model, CP violations are permitted in both the electroweak and strong nuclear sectors. In particular, the so-called $\Theta-$term of QCD, given by $S_{\Theta} = (\Theta/32\pi^2) \int  d^4x \, G^{\mu \nu}_a  \widetilde{G}_{a \,\mu \nu}$, where $G^{\mu \nu}_a$  and $\widetilde{G}^{\mu \nu}_a$ stand for the gluon field strength tensor and its dual, can generate a contribution to CP-violating photonic interaction \cite{Faccioli_PRD09}. Indeed, by assuming that $ \Theta \lesssim 10^{-10}$, which is consistent with the theoretical and experimental values for the neutron's electric dipole moment \cite{PDG}
($d_n \approx 5,2 \, \times \Theta \times 10^{-16} \, e \, \textrm{cm}$ and $d_n^{\textrm{exp}} \lesssim 10^{-26} \, e \, \textrm{cm}$),
the authors arrived at $\xi_3^{\textrm{QCD}} \lesssim 3,8 \times 10^{-47} \, T^{-2}$, which is at least 23 orders of magnitude smaller than the values of Euler-Heisenberg parameters.

Now we turn our attention to new physics beyond the Standard Model. According to Refs. \cite{Yamashita_PETP17,Ghasemkhani_EPJ23}, it is possible to generate a CP-violation Euler-Heisenberg-like electrodynamics by considering the inclusion of both vector and pseudo-vector (axial) couplings between fermions $(\psi)$ and photon $(A_\mu)$, namely, with the Lagrangian density interaction
\begin{equation} \label{L_va}
    L^{v,a} = - \bar{\psi} \, \gamma^\mu   \left( 
    g_v + g_a \gamma_5 \right) \psi \, A_\mu \, , 
\end{equation}
where $g_v$ and $g_a$ denote the vector and axial coupling constants, respectively. The correspondent effective field theory yields Eq. \eqref{L_CP} with the following parameters
\begin{eqnarray}
    \xi_1^{v,a} &=& \frac{1}{32 \, \pi^2 } \, 
    \frac{\hbar}{m^4 \, \varepsilon_0 \, c^5} \, \left[ 
    \frac{16}{45} \, g_v^4 + \frac{32}{15} \, g_v^2 \, g^2_a + \frac{16}{45} \, g_a^4 \right] \, , \label{xi_1_va} \\
    \xi_2^{v,a} &=& \frac{1}{32 \, \pi^2 } \, 
    \frac{\hbar}{m^4 \, \varepsilon_0 \, c^5} \, \left[ 
    \frac{28}{45} \, g_v^4 + \frac{56}{15} \, g_v^2 \, g^2_a + \frac{28}{45} \, g_a^4 \right] \, , \label{xi_2_va} \\
    \xi_3^{v,a} &=&  \frac{i}{48 \, \pi^2 } \, 
    \frac{\hbar}{m^4 \, \varepsilon_0 \, c^5} \,\left[ 
    g_v^3 \, g_a + g_v \, g_a^3 \right] \, . \label{xi_3_va} 
\end{eqnarray}
where $m$ corresponds to the fermion mass. For $g_a = 0$ and $g_v \equiv e$, we arrive at equivalent expressions for the Euler-Heisenberg parameters \eqref{EH_coeff}. At this point, we call attention to the reader that we have used the SI units and different conventions for the invariants $\mathcal{F}$ and $\mathcal{G}$, thereby the parameters \eqref{xi_1_va}$-$\eqref{xi_3_va} were adjusted correctly. Furthermore, we emphasize that the idea of calculating the effective theory from the interactions \eqref{L_va} was initially developed in the work \cite{Yamashita_PETP17} and, some years later, corrected in Ref. \cite{Ghasemkhani_EPJ23}. We also highlight that the new interaction in Eq. \eqref{L_va} should be interpreted as an effective contribution and not a fundamental one. For more details, we point out the Ref. \cite{Ghasemkhani_EPJ22}.


Similarly, one can consider non-minimal electromagnetic tensor (T) and pseudo-tensor (PT) couplings, such as 
\begin{equation} \label{L_P_PT}
    L^{n-m} =  \bar{\psi} \, \Sigma^{\mu \nu} \left( 
    g_T + i g_{PT} \, \gamma_5 \right) \psi \, F_{\mu \nu} \, , 
\end{equation}
where $\Sigma^{\mu \nu} \equiv \frac{i}{4} \left[ \gamma^\mu , \gamma^\nu \right]$, while $g_T$ and $g_{PT}$ denote the coupling constants. The correspondent effective field theory also generates the CP-violating model \eqref{L_CP}, but we have additional features here, namely, the couplings constants $g_T$ and $g_{PT}$ are related to the magnetic and electric dipole moments (MDM and EDM), respectively. For the electron case, the MDM measurement is in good agreement with the Standard Model predictions, while the EDM is predicted to be non-zero with a small value, being generated only by quantum corrections in the Standard Model and allowing a possibility for new physics \cite{Chupp_RMP19,PDG,Gorghetto_JHEP21}.

It is quite common to refer the model \eqref{L_CP} as CP-violating Euler-Heisenberg electrodynamics,  since we only have quadratic corrections in $\mathcal{F}$ and $\mathcal{G}$, as well as similar origins coming from effective theory involving photon and fermion interactions. However, it is also possible to obtain this model from other viewpoints. For instance, by including the axion field $\phi$ (a dark matter candidate) coupled to photon. The correspondent Lagrangian density is given by
\begin{equation} \label{L_phi}
    L_\phi = \frac{1}{2} \left( \partial_\mu \phi\right)^2 -\frac{1}{2} \, \frac{m^2_\phi \, c^2}{\hbar^2} \, \phi^2 + \frac{\widetilde{g}}{4} \, \phi \, F_{\mu \nu} \widetilde{F}^{\mu \nu} + 
    \frac{g}{4} \, \phi \, F_{\mu \nu} F^{\mu \nu} \, ,
\end{equation}
with $m_\phi$ being the axion mass. In addition, $g$ and $\widetilde{g}$ stand for the coupling constants. We highlight that is important to have both $g$ and $\widetilde{g}$ non-zero, which explicitly guarantees the breaking of CP symmetry, regardless of whether the axion is a scalar or pseudo-scalar field. By integrating out the axion field (see \cite{Gorghetto_JHEP21} and references therein), one arrives at the model \eqref{L_CP}, with the parameters
\begin{equation} \label{xis_axion}
    \xi_1^\phi = \frac{\mu_0}{2}  \frac{\hbar^2 g^2}{m_\phi^2 c^2} \, \; , \, \;
    \xi_2^\phi = \frac{\mu_0}{2} \frac{\hbar^2 \widetilde{g}^2}{m_\phi^2 c^2} \, \; , \, \;
    \xi_3^\phi = \mu_0 \frac{\hbar^2 \widetilde{g} g}{m_\phi^2 c^2} \, .
\end{equation}
Here, we clearly see that $g$ and $\widetilde{g}$ non-zero provide a CP-violating term $(\xi^\phi_3 \neq 0)$. The appearance of these interactions can be motivated in scenarios of axion-like particles (ALPs) coupled to the SM, which give rise to an effective chiral Lagrangian and electric dipole moments \cite{Paradisi_PRD21,Paradisi_JHEP24}. Moreover, the impact of light propagation in an external magnetic field with both $g$ and $\widetilde{g}$ couplings was investigated in Ref. \cite{Liao_PLB07}. It is worth mentioning that the idea of integrating out the axion field was initially developed in the works \cite{Bogorad_PRL19,Evans_PLB19,Xue_20} with $g=0$ and, posteriorly, generalized in Ref. \cite{Gorghetto_JHEP21}. In this manner, the effects of virtual axion-like particles generated contributions to non-linear electrodynamics. On the other hand, we would like to comment that there is another viewpoint in which the axion field is direct coupled to non-linear electrodynamics and contribute to a dispersive behavior of the electromagnetic waves in the presence of background fields \cite{Paixao_JHEP22,Paixao_JHEP24}.


Another dark matter model is discussed in Ref. \cite{Fan_PETP2018}, where the authors included a new abelian sector $U_{Y'}(1)$ coupled to the Electroweak sector $SU_L(2) \times U_Y(1)$ of the Standard Model. This dark sector has scalar, spinor and vector fields. The correspondent low-energy effective theory also yields the CP-violating model \eqref{L_CP} with more involved expressions for the parameters $\xi_1, \xi_2$ and $\xi_3$. Thus, contributions from dark matter proposals can also generate CP-violating non-linear electrodynamics.

As reported by Ref. \cite{Remmen_JHEP19}, some effective theories of the Standard Model, such as the inclusion of quantum corrections, can give rise to anomalous couplings. For instance, non-linear corrections in the hypercharge sector $U_Y(1)$ can produce anomalous neutral couplings between photon and $Z^0$ boson, as well as CP-violating  photon self-interactions. 

 We also highlight a recent work in which the inclusion of local Lorentz-violating backgrounds in the effective action leads to an extended Euler-Heisenberg-like model that encompasses the CP-violating contributions \cite{Wagno_PRD_2026}.  In addition, the introduction of both vector and pseudovector fields coupled to fermionic one can also lead to an effective action with terms that violate CP symmetry \cite{deSouza_EPCJ_2026}. Keeping these motivations in mind, some phenomenological investigations of the model \eqref{L_CP} can be applied in different contexts.


In principle, one could consider higher-order contributions in the expansion \eqref{gen_L_eff} such that the CP-violating terms $\mathcal{G} \mathcal{F}^2 $ and $\mathcal{G}^3$. Although these terms bring parameters $c_1, d_1, d_2$ and $d_3$ with dependencies on the magnetic background field, by dimensional analysis, the correspondent cut-offs will be larger than usual for $\mathcal{F}^2 \, , \mathcal{G}^2$ and $\mathcal{F G}$. For the purpose of investigating low-energy effects of CP-violating non-linear electrodynamics, the model \eqref{L_CP} is sufficient. 


On the other hand, one may adopt the point of view of a fundamental theory, as occurred in the case of the Born-Infeld electrodynamics \cite{Born_Infeld}, which was re-obtained from the context of String Theories \cite{Fradkin_PLB85,Bergshoeff_PLB87}. It is also worth mentioning that discrete symmetries, such as CPT, can be violated in String Theory scenarios and may produce effects in low-energy regime \cite{Colladay_PAN00}. Bearing these ideas in mind, let us propose a generalization of Born-Infeld electrodynamics described by
\begin{equation} \label{L_CP_BI}
    L_{nl}^{CP-BI} = \frac{ \Lambda^2}{c^2} \, \left[ 
    1 - \sqrt{1-\frac{2 c^2}{ \Lambda^2} \mathcal{F} 
    - \frac{c^4}{ \Lambda^4} \mathcal{G}^2 
    - \zeta \, \frac{c^4}{ \Lambda^4} \mathcal{F G}  } 
    \, \right] \, ,
\end{equation}
where $ \Lambda$ stands for the Born-Infeld parameter with the same dimension as the electric field. In addition, $\zeta$ is a dimensionless parameter associated with the CP-violating term. For $\zeta=0$, we recover the usual Born-Infeld (BI) model.
In the low-energy regime and weak fields, it is possible to show that the above CP-violating BI electrodynamics leads to
\begin{equation} \label{L_CP_BI_effe}
    L_{nl}^{CP-BI} \approx \mathcal{F} 
    + \frac{c^2}{2  \Lambda^2} \left[ \mathcal{F}^2
     + \mathcal{G}^2 + \zeta \, \mathcal{F G} \right] \, ,
\end{equation}
which assumes a similar form of Eq. \eqref{L_CP} with $\xi_1 = \xi_2 = c^2/(2 \Lambda^2)$ and $\xi_3 = \zeta \, c^2/(2\Lambda^2)$. To our knowledge, the non-linear electrodynamics \eqref{L_CP_BI} has not yet been investigated in the literature. Nevertheless, it is particularly important to note that extensions of BI electrodynamics have been proposed over the last three decades (see, for example, the papers \cite{Haas_POP23,Kruglov_JPA10,Gaete_EPJC14,Dehghani_EPJP22} and references contained therein). Therefore, with the aforementioned reasons and motivations, we shall concentrate our efforts on the phenomenological CP-violating model, given by Eq. \eqref{L_CP}, which encompasses most of effective field theories and models in low-energy regime.

Let us also mention here some possible values for the BI parameter. By adopting the original BI proposal, we have $ \Lambda_{BI} = 1,18 \times 10^{20} \, \textrm{V/m}$, which is associated with the maximum electric field that produces a finite electric self-energy equal to the electron rest energy $m_e c^2$. In this situation, the electric field of a point-like particle is finite at the origin. This value is two orders of magnitude greater than the critical Schwinger-Sauter electric field, given by $E_c \equiv m_e^2 c^3/(e\hbar) \approx 1,32 \times 10^{18} \, \textrm{V/m}$. In this manner, one can easily check that $\xi_1^{BI} = \xi_2^{BI} \equiv c^2/(2  \Lambda_{BI}^2) \approx 3,24 \times 10^{-24} \, T^{-2}$. We call attention to the fact that this value is of the same order of magnitude as the Euler-Heisenberg parameters $\xi_1^{EH}$ and $\xi_2^{EH}$. On the other hand, it is possible to adopt a bound derived from a high energy physics, namely, the analysis of the LHC measurement of light-by-light (LbyL) scattering, which takes into account the QED and SM predictions, provides a stronger limit $ \Lambda_{\textrm{LbyL}} \gtrsim 1,53 \times 10^{28} \, \textrm{V/m}$ \cite{Ellis_PRL17}, such that $\xi_1^{\textrm{BI-LbyL}} = \xi_2^{\textrm{BI-LbyL}} \equiv c^2/(2  \Lambda_{\textrm{LbyL}}^2) \lesssim  1,92 \times 10^{-40} \, T^{-2}$. In passing, it is opportune to comment that the Born-Infeld extension to the hypercharge sector $U_Y(1)$ also exhibits a similar bound as $ \Lambda_{\textrm{LbyL}}$, which can be obtained from investigations of anomalous gauge neutral couplings between  photon and  $Z^0$ boson \cite{Neves_EPJC22}.

In what follows, we consider the CP-violating model \eqref{L_CP}. Bearing in mind the coefficients \eqref{coef_c_d_LCP}
and assuming that $\xi_i \, {\bf B}_0^2 < 1 \, \; (i=1,2,3)$, it is possible to show that the dispersion relations \eqref{DR} simplify to
\begin{equation} \label{DRs_L_CP}
    \omega_{(\pm)} = c \, \sqrt{ k^2 - \Delta_{(\pm)} \, ( {\bf B}_0 \times {\bf k})^2} \, ,
\end{equation}
where we defined 
\begin{equation} \label{Delta_def}
    \Delta_{(\pm)} \equiv \xi_1 + \xi_2 \mp \sqrt{(\xi_1 - \xi_2)^2 + \xi_3^2} \, \; .
\end{equation}

Similarly, the correspondent group velocities are given by
\begin{equation} \label{vg_L_CP}
    {\bf v}_g \bigg|_{w_{(\pm)}}=  \frac{ c\,  {\bf \hat{k}} + c \, \Delta_{(\pm)} \, 
    [{\bf B}_0 \times ( {\bf B}_0 \times {\bf \hat{k}})] }{\sqrt{1 - \Delta_{(\pm)}  \, ( {\bf B}_0 \times {\bf \hat{k}})^2}} \, .
\end{equation}
After some algebraic manipulations, one can show that
\begin{equation} \label{vg_square}
   {\bf v}_g^2  =   c^2 - c^2 \, \left[
   \frac{\Delta_{(\pm)} ( {\bf B}_0 \times {\bf \hat{k}})^2 [1 - \Delta_{(\pm)}  {\bf B}_0^2 ]}{1-\Delta_{(\pm)} ( {\bf B}_0 \times {\bf \hat{k}})^2} \right] \, .
\end{equation}
In other words, by imposing the causality condition ${\bf v}_g^2 \leq c^2$, we find out that $\Delta_{(\pm)} \geq 0$, or equivalently, from the definition \eqref{Delta_def} we obtain
\begin{equation} \label{causal_cond}
    \xi_3^2 \leq 4 \, \xi_1 \xi_2 \, .
\end{equation}
Hence, with the parameters $\xi_1 $ and $ \xi_2$, one can arrive at a bound on $\xi_3$. A similar relation appears by investigating the analytic and unitary S-matrix of effective causal theory \cite{Remmen_JHEP19}.

At this point, it is instructive to check the consistency of our results with some investigations in the literature. Let us take the particular case with ${\bf B}_0 = B_0 \, {\bf \hat{x}}$ and ${\bf k} = k \, {\bf \hat{z}}$.  By using the dimensionless parameters $\chi_1 \equiv 2 \, \xi_1 B_0^2$, $\chi_2 \equiv 2 \, \xi_2 B_0^2$, $\chi_3 \equiv  \xi_3 B_0^2$ and assuming that $\chi_i < 1$,  the dispersion relation \eqref{DRs_L_CP} leads to
%
%
%
%
\begin{equation} \label{DR_hu_liao}
    \omega_{(\pm)} = ck \, \sqrt{1 -\frac{1}{2} (\chi_1 + \chi_2) \pm \sqrt{\frac{1}{4} (\chi_1 - \chi_2 )^2 + \chi_3^2}} \, ,
\end{equation}
which recovers the result in Ref. \cite{Liao_EPJ08}. Moreover, remembering the definition of the refractive index, $n_{(\pm)} \equiv c/v_p = ck/\omega_{(\pm)}$, one arrives at
\begin{equation} \label{ref_index_hu_liao}
    n_{(\pm)} \approx 1 + \frac{1}{4} (\chi_1 + \chi_2) \, \mp \, \frac{1}{2} \, \sqrt{\frac{1}{4} (\chi_1 - \chi_2)^2 + \chi_3^2} \, ,
\end{equation}
which also reproduces the result of Ref. \cite{Gorghetto_JHEP21}. 


Finally, we would like to present the current restrictions on the parameters $\xi_1$, $\xi_2$ and $\xi_3$ in SI units. The results were obtained and adapted from Refs. \cite{Akmansoy_PRD19,Gorghetto_JHEP21}. Essentially, we have two types of constraints related to low and high energy regimes. Let us first consider the low energy regime $( \hbar \, \omega < m_e c^2)$, where the effective photonic CP-violating model can be applied as a good approximation. The parameter $\xi_1$ can induce a correction to the Coulomb potential such that  investigations of energy levels of the hydrogen atom leads to $|\xi_1| \lesssim  1,05 \times 10^{-22} \, T^{-2}$. In addition, the following constraint  $ -1,38 \times 10^{-22} \, T^{-2} \lesssim \xi_2 \lesssim 1,75 \times 10^{-22} \, T^{-2}$ 
can be obtained by combining the previous bound on $\xi_1$ with the results of the PVLAS experiment \cite{PVLAS_PR20}, which measures the vacuum magnetic birefringence and induced ellipticity of polarized light passing through a region with a  magnetic background field. Bearing these results and the condition \eqref{causal_cond} in mind, we promptly arrive at $|\xi_3| \lesssim  2,79 \times 10^{-22} \, T^{-2}$. Therefore, we observe that, in the low energy regime, the parameters $\xi_i \, \;(i=1,2,3)$ are two orders above of the $\xi^{EH}$ and $\xi^{BI}$ parameters. On the other hand, in the high energy regime, we need to take into account the QED and SM predictions, which implies quite severe restrictions on the parameters associated with new physics Beyond the Standard Model (BSM), namely, based on the analysis of light-by-light scattering measurements of the LHC, it is possible to show that $\xi_i^{\textrm{BSM}} \lesssim 3,8 \times 10^{-42} \, T^{-2} $.   
These values are $20$ orders of magnitude lower than the limits obtained in the low energy regime.

 \section{Blackbody radiation and thermodynamic properties} \label{sec_3}

We are now ready to discuss the consequences of CP-violating non-linear electrodynamics on blackbody radiation thermodynamic laws, using a procedure analogous to Ref. \cite{Turcati_PRD23}. In this vein, we investigate the implications to thermodynamic parameters such as density energy, pressure, entropy, and specific heat. In addition, features related to deviations from the Planck distribution, Rayleigh-Jeans law, as well as the Wien displacement  and Stefan-Boltzmann laws will be discussed. Here, it is opportune to draw the reader's attention to our notation: in the parameters $\xi_i \, (i=1,2,3) $ and the magnetic background field ${\bf B}_0$ we have used the SI units, such that the correspondent dimensions are given by $[\xi_i] = T^{-2}$ and $[{\bf B}_0] = T$, with $T$ being the Tesla unit. However, for the thermodynamic quantities discussed in this Section, $T$ will denote the temperature in Kelvin (K) unit.

\subsection{The partition function and the thermodynamic properties}\label{partitionfunction}

To analyze blackbody radiation in the context of statistical mechanics, we consider the regime of non-zero temperatures well below the electron rest mass $m_{e}$, i.e., $k_{B}T\ll{m_{e}c^{2}}$, which ensures us that the photon sector gives the only contribution to the mentioned radiation \cite{Barton:1990mu,Kong:1998ic},  with $k_{B}$ and $T$ being the Boltzmann constant and the temperature, respectively. In other words, the photon gas is free of electron-positron pair production. Furthermore, since the modified dispersion relation does not change the foundations of statistical mechanics, we can use the grand-canonical formalism with a null chemical potential for the bosonic gas \cite{NiauAkmansoy:2013sxs,Anacleto:2018wlj}.

Let us then consider a system of $N$ photons with energy $\{E_{j}=\hbar \, {\omega_{j}}\}$, with $\omega_{j}$ being the angular frequency, where $j$ denotes $n_{j}$ particles with energy $E_{j}$ in each state. Additionally, the photon gas satisfies Bose-Einstein statistics. Therefore, the partition function is obtained from the sum over states, which takes the following form
\begin{eqnarray}
\mathcal{Z}=\sum_{n_{j}}\prod_{j}\left[e^{-\beta  \hbar {\omega}_{j}}\right]^{n_{j}} \, ,
\end{eqnarray}
where we adopted the usual notation $\beta=1/k_{B}T$.


In turn, the Helmholtz free energy is defined as
\begin{eqnarray}\label{helmholtz}
F=-\frac{1}{\beta} \ln \mathcal{Z},    
\end{eqnarray}
where 
\begin{eqnarray}
\ln \mathcal{Z}=-\sum_{i}ln\left(1-e^{-\beta  \hbar \omega}\right).    
\end{eqnarray}


The average energy in this situation reads

\begin{eqnarray}\label{ae}
\langle{E}\rangle=\sum_{{i}}\frac{\hbar{\omega}_{i}}{\left(e^{\beta\hbar{\omega}_{i}}-1\right)}.    
\end{eqnarray}

For large volume, the energy levels are very close to each other, which allows us to consider the energy as a continuous variable. In this limit, we can assume $\sum_{i}\rightarrow{(2\pi)^{-3}\int{d^{3}\mathbf{x}}\int{d^{3}\mathbf{k}}}$, and the relation (\ref{ae}) yields
\begin{eqnarray}
\langle{E}\rangle=\frac{1}{\left(2\pi\right)^{3}}\int{d^{3}\mathbf{x}}\int{d^{3}\mathbf{k}}\frac{\hbar{\omega}}{\left(e^{\beta\hbar{\omega}}-1\right)}.    
\end{eqnarray}

We are presently concerned with CP-violating non-linear contributions to the Planck distribution. Then, we need to make a variable change from the phase space to the frequency space, i.e., from $k$ to $\omega-$space.  Whenever there is anisotropic wave propagation, the relation between $\omega$ and $k$ variables is  given by the relation
\begin{eqnarray}
d\omega=\left(\mathbf{\nabla}_{k} \, \omega \right)\cdot{d\mathbf{k}},
\end{eqnarray}
where $\omega = \omega(k,\theta,\phi)$ for the most general case and
\begin{eqnarray}
\mathbf{v}_{g} \equiv \mathbf{\nabla}_{k} \, \omega = \mathbf{\hat{k}}\frac{\partial{\omega}}{\partial{k}}+\frac{\mathbf{\hat{\theta}}}{k}\frac{\partial{\omega}}{\partial{\theta}}+\frac{\mathbf{\hat{\phi}}}{k\sin \theta}\frac{\partial{\omega}}{\partial{\phi}}    \, ,
\end{eqnarray}
which denotes the group velocity.
%
%

The volume element transformation from the $k-$space to the frequency $\omega-$space depends on the modulus of the Jacobian transformation factor $J$, which leads to
\begin{eqnarray} \label{Jacobian}
J=\frac{\partial{k(w,\theta,\phi)}}{\partial{w(k,\theta,\phi)}}=\left(\frac{\partial{w}}{\partial{k}}\right)^{-1}.    
\end{eqnarray}

From the above relation, we note that only the radial component of the group velocity contributes to the transformation between $k$ and $w$ spaces (for more details and an alternative derivation, see the Appendix \ref{AppeendixA}).


Bearing this discussion in mind, we can use the result in eq. \eqref{vg_iso} and the corresponding phase velocity of the frequencies \eqref{DRs_L_CP} in the phase space transformation, such that 
%
\begin{eqnarray}
\langle{E}\rangle=\frac{V}{4\pi^{2}c^{3}\beta}\int^{\pi}_{0} {d\theta}\sin\theta\int_{0}^{\infty}d\omega \, {\omega}^{2}\frac{\hbar \, {\omega}}{\left(e^{\beta\hbar{\omega}}-1\right)}\Sigma_{(\pm)}\left(B_{0},\theta\right),
\end{eqnarray}
where we defined
\begin{eqnarray}\label{Sigma}
\Sigma_{(\pm)}\left(B_{0},\theta\right)=\frac{1}{\left(1-\Delta_{(-)}B_{0}^{2}\sin^{2}\theta\right)^{3/2}}+\frac{1}{\left(1-\Delta_{(+)}B_{0}^{2}\sin^{2}\theta\right)^{3/2}}.    
\end{eqnarray}

Here, it is important to emphasize that this factor $\Sigma_{(\pm)}$ incorporates the effects of non-linear electrodynamics, as well as the contribution from the magnetic background field ${\bf B}_0$ and the anisotropy related to the $\theta-$dependence.

Now, by performing the angular integral, we obtain 
\begin{eqnarray}\label{EEV}
\langle{E}\rangle=\frac{V}{\pi^{2}c^{3}\beta}\int^{\infty}_{0}{d}\omega \, \omega^{2} \, \Phi_{(\pm)}\left(\xi_{i},B_{0}\right)\frac{\hbar \,{\omega}}{\left(e^{\beta\hbar{\omega}}-1\right)}    ,
\end{eqnarray}
where we adopted the shorthand notation
\begin{eqnarray} \label{phi_pm}
{\Phi_{(\pm)}\left(\xi_{i},B_{0}\right) }= \frac{1-\left(\xi_{1}+\xi_{2}\right)B_{0}^{2}}{\left(1-\Delta_{(-)}B_{0}^{2}\right)\left(1-\Delta_{(+)}B_{0}^{2}\right)},    
\end{eqnarray}
and we recall again that $\xi_{i}$ with $i=1,2,3$ denote the parameters $\xi_{1},\xi_{2}$ and $\xi_{3}$. In addition, the CP-violating parameter $\xi_3$ appears in the definition of $\Delta_{(\pm)}$, given by Eq. \eqref{Delta_def}.

Concerning the energy density $u \equiv \langle{E}\rangle/V$, by integrating Eq. (\ref{EEV}) over all frequencies, we arrive at 
\begin{eqnarray}\label{totalenergydensity}
u\left(T\right)=aT^{4}, 
\end{eqnarray}
with the definition
\begin{eqnarray}\label{lettera}
a=\frac{4}{c}\left(\frac{\pi^{2}k_{B}^{4}}{15\hbar^{3}c^{2}}\right)\Phi_{(\pm)},
\end{eqnarray}
being the effective coefficient that encodes the CP-violating non-linear modifications and $\Phi_{(\pm)} \equiv \Phi_{(\pm)}\left(\xi_{i},B_{0}\right)$. 

From the above relations, the CP-violating non-linear corrections depend on $\Phi_{(\pm)}$, which have some interesting properties. First of all, when $B_0\rightarrow0$ or $\xi_{i}\rightarrow0$, we recover the usual result of Maxwell electrodynamics. Moreover, in the limit in which $\xi_{3}\rightarrow0$, the CP symmetry is preserved and the standard outcomes from CP-preserving non-linear electrodynamics are re-assessed \cite{Turcati_PRD23}.

At this stage, we shall compute the thermodynamic quantities. Accordingly to the relation (\ref{helmholtz}),
with the continuum limit $\sum_{i}\rightarrow{(2\pi)^{-3}\int{d^{3}\mathbf{x}}\int{d^{3}\mathbf{k}}}$ and the same phase space transformation, the free energy $F$ assumes the following form 
\begin{eqnarray}\label{freeenergy}
F=-V\left(\frac{\pi^{2}k_{B}^{4}{T}^{4}}{45\hbar^{3}c^{3}}\right)\Phi_{(\pm)}.
\end{eqnarray}

In this manner, the radiation pressure $P=-\left(\partial{F/\partial{V}}\right)_{T}$, the entropy $S=-\left(\partial{F/\partial{T}}\right)_{V}$, and the heat capacity $C_{V}=-\left(\partial{E/\partial{T}}\right)_{V}$ at constant volume densities are given by
\begin{eqnarray}\label{pressure}
P=\frac{a}{3}T^{4}, \quad S=\frac{4}{3}aT^{3}, \quad C_{V}=4aT^{3},
\end{eqnarray}
where we recall that the coefficient $a$ was defined in Eq. (\ref{lettera}). 

Regarding the equation of state between the energy density and the radiation pressure, one can promptly see that 
\begin{eqnarray}\label{eos}
P=\frac{u}{3},    
\end{eqnarray}
which remains unchanged compared to the standard results of Maxwell electrodynamics for the photon gas. Thus, the equation of state (\ref{eos}) is independent of the specific non-linear electrodynamics model under consideration, even if we consider CP-violating terms.
 We emphasize that this result was already expected due to the fact that the quanta involved are bosons with a gapless linear dispersion relation \eqref{DRs_L_CP}.

\subsection{Corrections to the Planck distribution}

Undoubtedly, the understanding of the blackbody radiation thermal laws led to the birth of quantum mechanics at the beginning of the last century. Thus, it is fair to turn our attention to describing the radiation properties of a blackbody in thermal equilibrium at temperature $T$. In this vein, an important quantity to use is the spectral density of energy $u$, per unit volume, which takes the following form
\begin{eqnarray}\label{spectralenergydensity}
u\left(\omega\right) d\omega =\left(\frac{{\omega}^{2}}{\pi^{2}c^{3}}\right)\frac{\hbar \, {\omega}}{\left(e^{\beta\hbar{\omega}}-1\right)} \, \Phi_{(\pm)} \, d\omega.
\end{eqnarray}

At this point, we shall highlight the effects of $\Phi_{(\pm)}-$term in the frequency spectrum. As is well-known, the density energy per unit of frequency can be cast as $u\left({\omega}\right)=g\left({\omega}\right) \epsilon\left({\omega}\right)$, where $\epsilon\left({\omega}\right)=\hbar {\omega} \, \left(1-e^{-\beta\hbar{\omega}}\right)^{-1}$ is the average energy per mode and 
\begin{eqnarray}\label{densityacessiblestates}
g\left({\omega}\right)=\frac{{\omega}^{2}}{\pi^{2}c^{3}}\Phi_{(\pm)} 
\end{eqnarray}
is the density of accessible states. The corrections to these results lead to an increase in the number of photons in each state, which shall become clearer in the next subsection, where we will present some graphics analyses.

At low frequencies, the distribution (\ref{spectralenergydensity}) reduces to
\begin{eqnarray}
u\left(\omega\right)=\left(\frac{{\omega}^{2}}{\pi^{2}c^{3}}\right)\left(k_{B}T\right)\Phi_{(\pm)},
\end{eqnarray}
which modifies the Rayleigh-Jeans law due to the magnetic background  field.

On the other hand, the energy density for each solid-angle element ($d\Omega=\sin\theta \, {d\theta} \, {d\phi}$) receives a contribution from the CP-violating term, which is given by
\begin{eqnarray}
u\left(T,\Omega\right)d\Omega =\left(\frac{\pi\kappa_{B}^{4}T^{4}}{120\hbar^{3}c^{3}}\right)\Sigma_{(\pm)}\left(B_{0},\theta\right)d\Omega.
\end{eqnarray}

To see more clearly the effect of the angular contribution for each solid-angle element, let us perturbatively expand the term $\Sigma_{(\pm)}$ in Eq. \eqref{Sigma}, which leads to
\begin{eqnarray}\label{SigmaApproach}
\Sigma_{(\pm)}\left(B_{0},\theta\right)\approx 2 +  \lambda_1 \,{\sin^{2}\theta}.    
\end{eqnarray}
with $ \lambda_1 \equiv 3(\xi_{1}+\xi_{2})B_{0}^{2}$. In this first-order perturbation, the CP-violating parameter $\xi_3$ does not contribute to $\Sigma_{(\pm)}$. Interestingly enough, the above relation shows us a monopole ($l=0$) and a quadrupole ($l=2$) contribution to the power angular spectrum. Furthermore, the maximal contribution will be in the direction perpendicular to the magnetic field. On the other hand, in the longitudinal direction, the non-linear component does not contribute. Observe that the temperature dependence remains proportional to $T^{4}$, where the deviations will be encoded in the intensity of the frequency spectrum.  Higher-order terms in the expansion will give further contributions to the power angular spectrum. For instance, by expanding Eq. (\ref{Sigma}) up to the fourth order, we promptly find
%
%
\begin{eqnarray}\label{Sigmafourth}
\Sigma_{(\pm)}\left(B_{0},\theta\right)\approx 2 +  \lambda_1 \, \sin^{2}\theta +  \lambda_2 \, \sin^{4}\theta,
\end{eqnarray}
where we defined
\begin{eqnarray}\label{beta}
 \lambda_2 \equiv\frac{15}{4}\left[2\left(\xi_{1}^{2}+\xi_{2}^{2}\right)+\xi_{3}^{2}\right]B_{0}^{4}.   
\end{eqnarray}

Therefore, we note that the contribution of $\xi_3$ appears as $\xi_3^2 B_0^4$, which results in an even smaller contribution to the energy density. Furthermore, besides the monopole and quadrupole terms, there is also the appearance of the eightfold component. 

Finally, we would like to stress that the Wien's law is not modified in our framework. As it is well-known, the Planck distribution for different surface temperatures peaks at a frequency that is directly proportional to the temperature. This is the so-called Wien's displacement law. The maximum occurs by finding the solution of $du(\omega)/d\omega=0$, which, in our case, remains the same as the Planck law, i.e, 
$\omega_{\max} \approx 2.822 \, {k}_{B} \, T/\hbar$.




\subsection{Radiance and the modified Stefan-Boltzmann law}

The Stefan-Boltzmann law has several applications in physics, including the observation of emitted radiation of astrophysical objects and cosmology, among others. Therefore, it would be interesting to evaluate the CP-violating non-linear contributions in the present scenario. 

The total radiance is defined by the total energy emitted per unit time and per unit area of the cavity surface, given by
\begin{eqnarray}\label{definition}
R\left(T\right)=\int_{0}^{2\pi}{d\phi}\int_{0}^{\pi/2}d\theta \,{\sin\theta}\cos\theta \int_{0}^{\infty}B\left(\omega,\theta,T\right) d \omega,
\end{eqnarray}
where the spectral radiance density is
\begin{eqnarray}
B\left(\omega,\theta,T\right)=\frac{\omega^{2}}{8\pi^{3}c^{2}}\left(\frac{\hbar{\omega}}{e^{\beta\hbar{\omega}}-1}\right)\Sigma_{(\pm)}\left(B_{0},\theta\right).    
\end{eqnarray}

Small deviations are expected from the standard results in the blackbody radiation laws, which allow us to use the relation (\ref{Sigmafourth}). Hence, it is straightforward to derive the generalized Stefan-Boltzmann law from the integral $(\ref{definition})$, which takes the following form
\begin{eqnarray}\label{stefanboltzmannlaw}
R\left(T\right)=\sigma_{eff}T^{4},
\end{eqnarray}
where we defined the effective Stefan-Boltzmann constant as
\begin{eqnarray}\label{effectivesigma}
\sigma_{eff}=\sigma\left[1+\frac{ \lambda_1}{4}+\frac{\lambda_2}{6}\right] \, ,
\end{eqnarray}
with $\sigma=\left(\pi^{2}k_{B}^{4}/60\hbar^{3}c^{2}\right)$ being the usual Stefan-Boltzmann constant. 

The lowest order-corrections in the effective Stefan-Boltzmann constant (\ref{effectivesigma}) depends on the parameters $\xi_1$, $\xi_2$, $\xi_3$, and the magnetic background field $B_0$ through the coefficients $\lambda_1$ and $\lambda_2$. We first note that whenever $ \lambda_1 \rightarrow 0$ and $\lambda_2 \rightarrow 0 $, we recover the usual Stefan-Boltzmann law. In addition, for $\lambda_2 \rightarrow 0$, one arrives at the well-known results of CP-preserving non-linear electrodynamics \cite{Turcati_PRD23}. Within the first-order perturbation with $ \lambda_1$, the CP-violating parameter is not relevant to the total radiance flux. However, taking into account the second-order perturbation due to $ \lambda_2$, there is the appearance of the $\xi_{3}-$correction. We also note that in our formalism, the connection between total radiance and energy density, mediated by the geometric factor $c/4\pi$, is lost due to non-linear effects, which lead to a model dependence, as one can see by comparing relations  (\ref{totalenergydensity}) and (\ref{stefanboltzmannlaw}). More specifically, the angular dependence on the spectral density modifies both the radiance and the energy density in such a way that the role of the $\xi_{i}-$coefficients cannot be neglected, in contrast to what happens in Maxwell theory, where a geometric factor interlinks these quantities. 

To conclude this subsection, we would like to point out that although the Stefan-Boltzmann constant is modified in this situation, the temperature dependence remains the same as the usual blackbody radiation results, i.e., depends only on $T^{4}$. Corrections in the temperature can be induced, for instance, by higher-order momenta or energy terms in the dispersion relation, which is not the case in our formalism. 


\subsection{Applications and phenomenological analyses} 

We are now in a position to evaluate the effects of the CP-violating non-linearity in the frequency spectrum by comparing our model with the Planck distribution. Specifically, we 
shall consider the values of the $\xi_{i}-$coefficients within the frameworks of the Euler-Heisenberg and Born-Infeld models as well as the most recent laboratory constraints on the non-linear parameters, and then investigate the contribution to the blackbody radiation spectrum. In addition, we would like to stress that the perturbative expansion (\ref{L_CP}) is valid even for strong magnetic fields, such as in the neighborhood of magnetars, which allow us to adopt magnetic fields above the Schwinger limit once condition $\xi_{i}{\bf{B}}_{0}^{2}<1$ is satisfied. The exception is the Euler-Heisenberg electrodynamics, where fields around the Schwinger limit give rise to electron-positron pairs, and the perturbative expansion is not valid anymore. For the Euler-Heisenberg case, we will consider fields of magnitude close to $10^7 \, T$, which ensures the validity of our approximation.

As discussed in Section \eqref{subsec_2C}, the Euler-Heisenberg parameters assume the values $\xi^{EH}_{1} \approx 5,28 \times 10^{-24} \, T^{-2}$ and $\xi^{EH}_{2} \approx 9,24 \times 10^{-24} \, T^{-2}$, while the Born-Infeld coefficients are both given by $\xi^{BI}_1 = \xi^{BI}_2 \approx3,24\times10^{-24} \, T^{-2}$. 

To estimate the deviations from the Planck distribution, one can examine
\begin{eqnarray}
\frac{\delta{u}}{u}=\frac{u_{M}-u}{u},    
\end{eqnarray}
where $u$ concerns the Planck distribution and $u_{M}$ regards the non-linear electrodynamics corrections to Planck's law, which gives us
\begin{eqnarray} \label{rel_dev}
\frac{\delta{u}}{u}=\Phi_{\pm}-1 \, ,   
\end{eqnarray}
where $\Phi_{(\pm)} \equiv \Phi_{(\pm)}\left(\xi_{i},B_{0}\right)$ is defined in Eq. \eqref{phi_pm}.

\begin{center}
\begin{figure}[htb]
\begin{minipage}{0.5\textwidth}
{\includegraphics[scale=0.65]{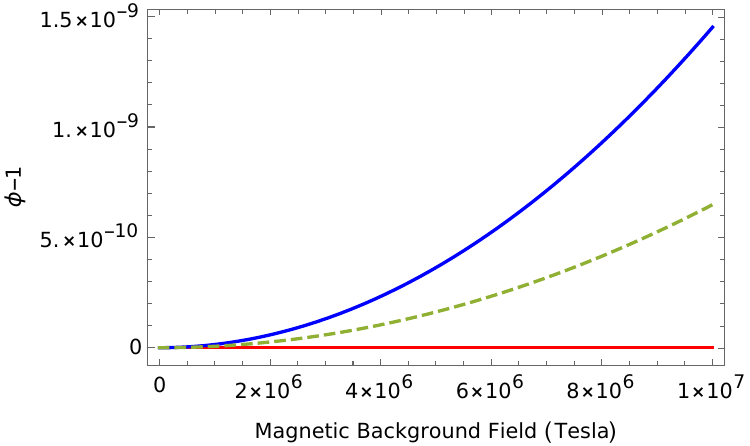}}
\end{minipage}%
\caption{Relative deviations in Eq. \eqref{rel_dev} of Euler-Heisenberg (blue line) and Born-Infeld (dashed green line) models from Planck's blackbody radiation law (red line). We have assumed a temperature of $T=6 \times 10^5 \,K$ and a magnetic background field $B_0=10^{7} \, T$.}\label{Fig1}
\end{figure}
\end{center}

Clearly, our framework is suitable for applications in astrophysics with an intense magnetic background field. In this sense, it is well-known that the thermal luminosity of neutron stars, inferred from observations, is mostly emitted in soft X-rays \cite{Morton1964}, with a superficial temperature estimated in $T\approx(10^{5}-10^{6})\,K$ \cite{Lattimer:2015eaa}. The surface magnetic field of these astrophysical objects, in turn, lies between $10^{5}\, T$ and $10^{10}\,{T}$ \cite{Olausen:2013bpa}. Therefore, taking a magnetic background field intensity of $B_{0}\approx10^{7}\,T$ and a temperature $T=6\times10^{5}\,K$, which clearly sa-tisfies conditions $k_B T << m_e c^2 $ and $\xi_i {\bf B}_0^2 < 1$, we obtain Fig. (\ref{Fig1}). This graph illustrates that the corrections arising from CP-preserving non-linearity lead to an increase in the number of accessible states at each frequency. Consequently, these findings are consistent with previous studies on a photon gas in thermal equilibrium at temperature T, within the formalism of non-linear electrodynamics models without CP-violating contributions \cite{Turcati_PRD23}.

\newpage
The effects of CP-violation in the context of blackbody radiation thermal laws can be examined by considering constraints on the $\xi_{i}-$coefficients arising from experiments at low energies. At this moment, we recall the results discussed in Section \eqref{subsec_2C}, where the combined constraints related to the small deviations from the shift levels of the hydrogen atom and PVLAS experiment lead to $|\xi_1| \lesssim 1,05 \times 10^{-22} \, T^{-2}$ and $-1,38 \times 10^{-22} \, T^{-2} \lesssim \xi_2 \lesssim 1,75 \times 10^{-22} \, T^{-2}$ such that, with the causality condition (\ref{causal_cond}), one can easily check that $|\xi_{3}| \lesssim 2.79 \times 10^{-22} \, T^{-2}$. In what follows, we adopt the upper bounds for $\xi_1$ and $\xi_2$, with some analyzes of different values for $\xi_3$.

\begin{center}
\begin{figure}[htb]
\begin{minipage}{0.5\textwidth}
{\includegraphics[scale=0.8]{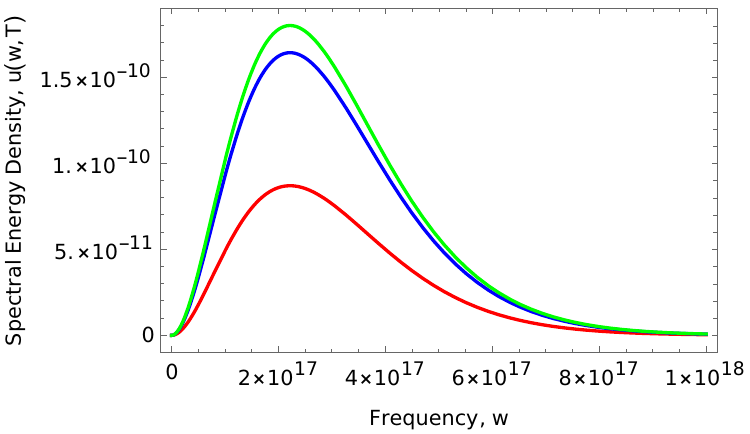}}
\end{minipage}%
\caption{Graph of the spectral energy density distribution for a surface temperature $T=6 \times 10^5 \,K$ and a magnetic background field $B_0= 4 \times 10^{11} \, T$. The red line corresponds to the Planck spectrum, while the blue and green are associated with CP-preserving and CP-violating non-linear electrodynamics, respectively.}\label{Fig3}
\end{figure}
\end{center}

A glance at Fig. (\ref{Fig3}) for a magnetic background field $ B_{0}= 4 \times10^{11} \, T$ above the critical value $B_{c}\approx10^{9}\,T$ and surface temperature $T=6 \times 10^5 \, K$ is enough to convince us that CP-violating non-linear corrections could, at least in principle, be observed for strong magnetic fields. The red line corresponds to the Planck spectrum, while the blue line stands for the CP-preserving non-linear electrodynamics $(\xi_3 =0)$. For the particular case with $\xi_3 = 10^{-22} \, T^{-2}$, we note that the inclusion of the CP-violating parameter (green line) leads to an increase of the peak. On the other hand, it is possible to show that, for the magnitude values $\xi_3 = 10^{-23} \, T^{-2}$ and below, there are no significant contributions, so the lines will remain very close to the CP-preserving situation. Indeed, when the CP-violating parameter is of the same magnitude order as $\xi_{1}$ and $\xi_{2}$, the CP violation is more prominent.

\begin{center}
\begin{figure}[htb]
\begin{minipage}{0.5\textwidth}
{\includegraphics[scale=0.65]{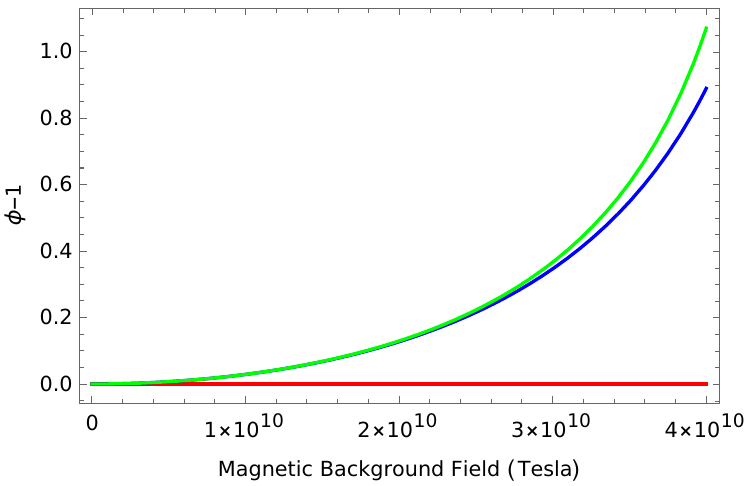}}
\end{minipage}%
\caption{Relative deviations in Eq. \eqref{rel_dev} of CP-preserving (blue line) and CP-violating (green line) models from Planck's blackbody radiation law (red line).}\label{Fig4}
\end{figure}
\end{center}

\newpage
Finally, we exhibit in Fig. (\ref{Fig4}), the plot of the relative deviations in Eq. \eqref{rel_dev} for CP-preserving and CP-violating non-linear electrodynamics associated with the blue and green lines, respectively. As before, the red line corresponds to the Plank's case. We now consider a magnetic background field of magnitude $B_{0}\approx10^{10} \,T$,  temperature $T=6\times10^{5}K$, and $\xi_3 = 10^{-22} \, T^{-2}$. 


In this situation, we note that the effect of CP-violating non-linear electrodynamics should be relevant for strong magnetic fields, namely, the deviation from CP-preserving case appears above $B_0 = 2 \times 10^{10} \, T$, when the blue and green lines turn out separated. Again, for the values of magnitude $\xi_3 = 10^{-23} \, T^{-2}$ and below, we did not observe relevant contributions.

Therefore, we conclude that the effects of non-linear electrodynamics and CP-violating terms become more relevant
in the presence of strong magnetic fields with a magnitude of $B_0 = 10^7 \, T$ and higher orders, which encompasses some astrophysical scenarios, such as neutron stars. Nevertheless, within the cosmic microwave background (CMB) scale, where the expected magnetic field is of the order $10^{-13} \, T$, we point out that there will be no relevant contributions to the power spectrum. Our phenomenological investigations with experimental constraints for $\xi_i$ indicate that the CP-violating contribution $\xi_3$ is non-trivial when it assumes the same order of magnitude as the CP-preserving parameters $\xi_1$ and $\xi_2$ with the causality condition $\xi_3^2 \leq 4 \, \xi_1 \xi_2 $ being respected.

\section{Concluding Comments and Perspectives} \label{sec_concl}

The purpose of this contribution has been to investigate non-linear electrodynamics \mbox{models} where the CP symmetry is not preserved. In this vein, we have considered in detail the field equations, wave-propagation solutions and group velocities in the presence of a magnetic background field ${\bf B}_0$. It should be highlighted that the existence of a non-trivial CP-violating term leads to a bi-anisotropic vacuum, i.e., both electric and magnetic perturbative fields can generate the polarization and magnetization tensors. This result contrasts with the usual anisotropic vacuum behavior in non-linear electrodynamics, where the electric field is responsible for the polarization tensor and the magnetic field for the magnetization tensor. Moreover, we have assumed the particular model \eqref{L_CP}, which encompasses most of the effective CP-violating proposals in the literature, described by the coefficients $\xi_{i} \, (i=1,2,3)$. By analyzing the correspondent group velocity ${\bf v}_g$ and imposing the causality condition $({\bf v}_g^2 \leq c^2)$, we arrive at the constraint $\xi_3^2 \leq 4 \, \xi_1 \xi_2 $, i.e., an upper bound for the CP-violating contribution in terms of the CP-preserving ones.

We have also investigated the consequences for photons at the temperature below the electron-positron production in a magnetized vacuum. We have then shown the appearance of small deviations in the thermodynamic quantities induced by the external magnetic field ${\bf B}_0$ and the coefficients $\xi_{i} $.  We have checked that the equation of state between the modified energy density and the radiation pressure remains unchanged, namely, $P=u/3$, even if we consider CP-violating terms $(\xi_3 \neq 0)$, being consistent with the fact that the dispersion relation  \eqref{DRs_L_CP} remains linear in $k = |{\bf k}|$.

Regarding blackbody radiation, we have found modifications of the Planck distribution and the Stefan-Boltzmann law arising in the scenario of the CP-violating non-linear electrodynamics. Specifically, the energy density is modified in such a way that there is an increase in the number of accessible states, leading to a more prominent curve in the frequency spectrum. Indeed, by using  phenomenological results for the coefficients $\xi_{i}$, the contributions of CP-preserving and CP-violating non-linear terms were identified in Figs. (\ref{Fig3}) and (\ref{Fig4}).  Our results indicate that the CP-violating contribution $\xi_3$ is non-trivial when it assumes the same order of magnitude as the CP-preserving ones, $\xi_1$ and $\xi_2$.


Finally, we would like to point out some perspectives.
Here, we have considered the modified dispersion relations $\omega_{\pm}$ to derive the partition function $\mathcal{Z}$, the average energy $\langle{E}\rangle$ and, consequently, to arrive at the modified thermodynamics quantities and the corresponding blackbody radiation thermal laws. However, one could adopt other point of view, by constructing the energy-momentum tensor of a photon gas and then apply the imaginary time formalism to derive the partition function \cite{Barton:1990mu}. The energy-momentum tensor of a single photon is well-known in the context of non-linear electrodynamics (see, for example, Refs. \cite{Paixao_JHEP22,Ospedal_RBrEF}). Nevertheless, the generalization to a photon gas is not trivial and an anisotropic pressure component could potentially arise.  In addition, to deal with the partition function, a linearization procedure will be needed in order to find analytical solutions. It remains to be investigated in a future work, as well as to carry out comparisons between the two methodologies.

Moreover, in regard to  observational data,  the Stefan-Boltzmann law can be linked to the luminosity of stars, where the surface temperature of a stellar object, such as the sun, is treated approximately as a blackbody. Since non-linear electrodynamics have been considered in this work, it would be interesting to compare, in this case, the frequency spectrum generated by astrophysical objects in a strong magnetic field medium. For instance, this is the case of magnetars, which are neutron stars with intense magnetic fields. Nowadays, there are many observational data sets from these objects. Their spectrum is mainly composed of a soft X-ray thermal component ($k_B \, T\approx0.5~$keV) and a non-thermal power law tail ($k_B \, T\approx100-200~$keV)  \cite{Turolla:2015mwa}, where \textrm{eV} denotes the electron-volt unit $( \textrm{keV} = 10^3 \textrm{eV})$. In this sense, as a further step, we hope that a detailed analysis of the Chandra X-ray Observatory, XMM-Newton, and Suzaku data could be useful to constrain the parameters of the effective non-linear electrodynamics \eqref{L_CP}. In particular, the Chandra X-ray Center offers an open-source software package for the analysis of the Chandra X-ray telescope data \cite{Fruscione2006}. The Chandra Interactive Analysis of Observations (CIAO) includes algorithms for source detections, spectroscopy, temporal analysis, and image processing. The implementation of our model in the CIAO software package could be very useful for imposing bounds on the parameters $\xi_{i}$ with these data and trying to observe the role of the CP violation in the electromagnetic sector. Similar methodology could be applied with data coming from XMM-Newton and Suzaku telescopes.


Last but not least, by assuming the possibility of parity violation at high energies, one might investigate if the CP-violating Born-Infeld model \eqref{L_CP_BI} could be generated from the low-energy limit of String Theories. It is also worthy to mention that we only consider the effects of CP-violating non-linear electrodynamics in vacuum. However, there are many motivations in the literature to examine non-linear electrodynamics in material media (see, for instance, some recent applications in Dirac materials \cite{Keser_PRL22,Neves_JPA23,Querts_PRA24}). Specifically, we highlight the case of chiral media \cite{Manoel_PRB24}, where the current density is proportional to the magnetic field and, consequently, the equations of motion in matter break parity symmetry. Investigations of non-linear electrodynamics with $\xi_3 \neq 0$ in chiral media and their optical effects have not  been carried out. We expect to report on these issues elsewhere.


\appendix

\section{Anisotropic surface parametrization}\label{AppeendixA}

The volume element in the $k-$space between $\mathbf{k}$ and $\mathbf{k}+d\mathbf{k}$ is given by
\begin{eqnarray}
d^{3}k=k^{2\,}\sin\theta\,dk\,d\theta\,d\phi.\end{eqnarray}

In a general media defined by the relation $\omega(\mathbf{k})$, the thermal radiation distribution
differs from the standard result of the Planck law due to anisotropic emission. Therefore, it is important to be careful in the transformation from $k-$space to $\omega-$space.

To achieve that, we need to parametrize an isosurface, i.e., express $d^{3}k=dS\,dn$, where $dS$ is the infinitesimal surface area for the constant$-\omega$ and $dn$ is the infinitesimal displacement perpendicular to this surface.


The parametrization of anisotropic isosurface in spherical coordinates is given by the equation
\begin{eqnarray}
\omega\left(\mathbf{k}\right)=\omega_{0},
\end{eqnarray}
where $\omega$ is a function of $\mathbf{k}=(k,\theta,\phi)$ and $\omega_{0}$ is a constant representing an isosurface. Assuming that we can solve $k$ in terms of $\theta$ and $\phi$, i.e., $k=k(\theta,\phi)$, we are left, locally, with a two-dimensional surface embedded in a three-dimensional one. Therefore, we can use the parameters $\theta$ and $\phi$ as surface parameters.

Under these assumptions, we have
\begin{eqnarray}
\mathbf{k}\left(\theta,\phi\right)=k(\omega_{0},\theta,\phi) \, \hat{\mathbf{k}}(\theta,\phi).    
\end{eqnarray}

We can then define two linearly independent vectors 
\begin{eqnarray}
\mathbf{t}_{\theta}=\frac{\partial{\mathbf{k}}}{\partial{\theta}} \,, \quad\quad \mathbf{t}_{\phi}=\frac{\partial{\mathbf{k}}}{\partial{\phi}} \,,    
\end{eqnarray}
which form a tangent plane to each point of the isosurface.

On the other hand, the orthonormal vector to the surface is given by the gradient of the $\omega-$function, namely,
\begin{eqnarray}\label{normalvector}
\mathbf{\hat{n}}=\frac{\nabla_{k} \,\omega}{|\nabla_{k} \,\omega|}    
\end{eqnarray}
where 
\begin{eqnarray}
\nabla_{k} \,\omega=\mathbf{\hat{k}}\frac{\partial\omega}{\partial{k}}+\frac{\mathbf{\hat{\theta}}}{k}\frac{\partial\omega}{\partial\theta}+\frac{\mathbf{\hat{\phi}}}{k \sin\theta}\frac{\partial\omega}{\partial\phi},    
\end{eqnarray}
is the group velocity. The modulus of the group velocity yields
\begin{eqnarray}
|\nabla_{k}\,\omega|&=&\sqrt{\left(\frac{\partial\omega}{\partial{k}}\right)^{2}+\frac{1}{k^{2}}\left(\frac{\partial\omega}{\partial\theta}\right)^{2}+\frac{1}{k^{2} \sin^{2}\theta}\left(\frac{\partial\omega}{\partial\phi}\right)^{2}}.
\end{eqnarray}

The differential $d\omega$, in turn, takes the form
\begin{eqnarray}\label{dw}
d\omega=\left(\nabla_{k}\,\omega\right)\cdot{d\mathbf{k}},    
\end{eqnarray}
where
\begin{eqnarray}\label{dkexpansion}
d\mathbf{k}=\frac{\partial\mathbf{k}}{\partial{\omega}}d{\omega}+\frac{\partial\mathbf{k}}{\partial{\theta}}d\theta+\frac{\partial\mathbf{k}}{\partial{\phi}}d\phi.    
\end{eqnarray}

Inserting relation (\ref{dkexpansion}) into (\ref{dw}), we find
\begin{eqnarray}\label{dwsimplified}
d\omega=\left(\nabla_{k} \,\omega\right)\cdot\frac{\partial\mathbf{k}}{\partial\omega}d\omega,
\end{eqnarray}
since, by construction, $\mathbf{t}_{\theta}$ and $\mathbf{t}_{\phi}$ are both orthogonal to the surface and gives null contribution. 

At this point we should ask ourselves what is the connection between the vetor $\mathbf{t}_{w}=\partial\mathbf{k}/\partial\omega$ and the normal vector $\mathbf{\hat{n}}$. To answer this question, we note that, by using relation (\ref{normalvector}) into (\ref{dwsimplified}), we find
\begin{eqnarray}
\mathbf{\hat{n}}\cdot{\frac{\partial\mathbf{k}}{\partial \omega}}=\left|\nabla_{k} \,\omega\right|^{-1}.
\end{eqnarray}

In addition, the $d\omega$ term can be cast as
\begin{eqnarray}
d \omega = \left(\nabla_{k}\,\omega\right)\cdot{d\mathbf{k}}=\left|\nabla_{k}\,\omega\right|\mathbf{\hat{n}}\cdot{d\mathbf{k}},    
\end{eqnarray}
where we can define $\mathbf{\hat{n}}\cdot{d}\mathbf{k}=dn$. Indeed, the term $\mathbf{\hat{n}}\cdot{d}\mathbf{k}$ is the orthogonal infinitesimal displacement of two frequency isosurfaces between $\omega$ and $\omega+d\omega$. In other words, is the projection of $d\mathbf{k}$ in the normal direction, which give us
\begin{eqnarray}
dn=\frac{d\omega}{\left|\nabla_{k}\,\omega\right|}.    
\end{eqnarray}

Therefore, the transformation from the $k-$space to the $\omega-$space assumes the form
\begin{eqnarray}\label{generalform}
d^{3}k=\frac{dw\,{d}S}{|\nabla_{k}\,\omega|}, 
\end{eqnarray}
with 
\begin{eqnarray}
dS=\left|\frac{\partial{\mathbf{k}}}{\partial{\theta}}\times\frac{\partial{\mathbf{k}}}{\partial{\phi}}\right|d\theta{d}\phi  \end{eqnarray}
being the infinitesimal surface area. 

The differential surface element, after some algebraic manipulations, leads to
\begin{eqnarray}\label{diffsurface}
dS=k^{2} \sin\theta\sqrt{1+\frac{1}{k^{2}}\left(\frac{\partial{k}}{\partial\theta}\right)^{2}+\frac{1}{k^{2} \sin^{2}\theta}\left(\frac{\partial{k}}{\partial\phi}\right)^{2}}d\theta{d\phi}.    
\end{eqnarray}

In a constant surface, we have 
\begin{eqnarray}
\frac{d\omega}{d\theta}=\frac{d\omega}{d\phi}=0,    
\end{eqnarray}
which give us the following relations:
\begin{eqnarray}\label{relation-1}
\frac{\partial{k}}{\partial\theta}=-\frac{\partial\omega/\partial\theta}{\partial\omega/\partial{k}} \, ,\\
\label{relation-2}
\frac{\partial{k}}{\partial\phi}=-\frac{\partial\omega/\partial\phi}{\partial\omega/\partial{k}} \, .
\end{eqnarray}

By applying the relations (\ref{relation-1}) and (\ref{relation-2}) in the differential surface element (\ref{diffsurface}) and rearranging the terms, one can promptly show that Eq. (\ref{generalform}) takes the form
\begin{eqnarray}
d^{3}k=\left|\frac{\partial\omega}{\partial{k}}\right|^{-1}k^{2}(\omega,\theta,\phi) \sin\theta \, {d\omega} \, d\theta \, {d\phi},    
\end{eqnarray}
in accordance with  the Jacobian transformation factor, Eq.
\eqref{Jacobian}, discussed in the Sec. (\ref{partitionfunction}). 

Whenever the surface reduces to $\omega(\mathbf{k})=ck$, one recovers the usual outcomes from the blackbody radiation thermal laws. We also emphasize that the transformation from the phase space to the frequency space is general and valid in any situations where there is an anisotropic distribution of radiation in thermal equilibrium. Note also that the mentioned transformation involves only the radial component of the group velocity, i.e., $\mathbf{\hat{k}}\cdot\mathbf{v}_{g}=\partial\omega/\partial{k}$.

Finally, it is worth mentioning that, in condensed matter systems \cite{Cole:1977}, where there is also anisotropic wave propagation, the density of accessible states takes the following expression
\begin{eqnarray}
g\left( \omega \right)=\frac{1}{(2\pi)^{3}}\int_{\omega}\frac{dS_{\omega} \, d\omega}{|\nabla_{k} \, \omega|},    
\end{eqnarray}
where the integral is performed around surfaces of $\omega-$constant. We remark that both approaches are equivalent.


{\bf Acknowledgments:}  The authors thank José A. Helayël-Neto, Dmitri P. Sorokin, and Rodrigo Bufalo  for helpful comments and exchange of ideas. L.P.R. Ospedal is grateful to the support of the {\it  Fundação Carlos Chagas Filho de Amparo à Pesquisa do Estado do Rio de Janeiro} (FAPERJ) for his post-doctoral senior fellowship. R. Turcati acknowledges the financial support from the $PCI$ program of the Brazilian agency {\it Conselho Nacional de Desenvolvimento Científico e Tecnológico}  (CNPq).  S.B. Duarte also thanks CNPq for partial financial support.


\end{document}